\documentclass[letterpaper,11pt]{article}
\usepackage{fullpage}
\usepackage{amsmath,amsthm}
\usepackage[numbers,sort]{natbib}

\usepackage{times}
\usepackage{soul}
\usepackage{url}
\usepackage[hidelinks]{hyperref}
\usepackage[utf8]{inputenc}
\usepackage[small]{caption}
\usepackage{graphicx}
\usepackage{amsmath}
\usepackage{amsthm}
\usepackage{booktabs}
\usepackage{textcomp}
\usepackage{color}
\usepackage{amsfonts,amssymb,bbm}
\usepackage{empheq}
\usepackage{cleveref}
\usepackage{xcolor}
\usepackage{tcolorbox}
\usepackage{tikz}
\usetikzlibrary{shadows,arrows.meta,positioning,backgrounds,fit,chains,scopes}
\usepackage{caption}
\usepackage{subcaption}

\tikzset{%
  materia/.style={draw, fill=blue!20, text width=5em, text centered, minimum height=1.5em},
  etape/.style={materia, text width=5em, minimum width=6em, minimum height=3em, rounded corners, drop shadow},
  line/.style={draw, thick, color=black!70, -LaTeX},
}

\usepackage[noend,ruled]{algorithm2e}

\SetCommentSty{mycommfont}
\SetKwInput{KwInput}{Input}
\SetKwInput{KwOutput}{Output}  
\SetAlFnt{\small}
\SetAlCapFnt{\small}
\SetAlCapNameFnt{\small}
\SetAlCapHSkip{0pt}
\IncMargin{-\parindent}

\newtheorem{theorem}{Theorem}

\newcommand{\argmax}{\mathrm{argmax}}

\newcommand{\one}{\mathbbm{1}}
\newcommand{\set}[1]{\left\{ #1 \right\}}
\newcommand{\abs}[1]{\left| #1 \right|}
\renewcommand{\bar}{\overline}

\newcommand{\calL}{\mathcal{L}}
\newcommand{\calP}{\mathcal{P}}

\newcount\Comments  
\newcommand{\kibitz}[2]{\ifnum\Comments=1{\color{#1}{#2}}\fi}

\newcommand{\prior}{\calP}
\newcommand{\signal}{\textstyle\Pr_s}
\newcommand{\posterior}{\textstyle\Pr_g}
\newcommand{\inference}{\textstyle\Pr_o}

\title{Surprisingly Popular Voting Recovers Rankings, Surprisingly!}

\author{
Hadi Hosseini\\
Pennsylvania State University\\
\texttt{hadi@psu.edu}
\and
Debmalya Mandal\\
Columbia University\\
\texttt{dm3557@columbia.edu}
\and
Nisarg Shah\\
University of Toronto\\
\texttt{nisarg@cs.toronto.edu}
\and
Kevin Shi\\
University of Toronto\\
\texttt{kevins.shi@mail.utoronto.ca}
}

\date{}

\begin{document}

\maketitle

\begin{abstract}
The wisdom of the crowd has long become the de facto approach for eliciting information from individuals or experts in order to predict the ground truth. However, classical democratic approaches for aggregating individual \emph{votes} only work when the opinion of the majority of the crowd is relatively accurate. A clever recent approach, \emph{surprisingly popular voting}, elicits additional information from the individuals, namely their \emph{prediction} of other individuals' votes, and provably recovers the ground truth even when experts are in minority. This approach works well when the goal is to pick the correct option from a small list, but when the goal is to recover a true ranking of the alternatives, a direct application of the approach requires eliciting too much information. We explore practical techniques for extending the surprisingly popular algorithm to ranked voting by partial votes and predictions and designing robust aggregation rules. We experimentally demonstrate that even a little prediction information helps surprisingly popular voting outperform classical approaches.
\end{abstract}


\section{Introduction}
The wisdom of the crowd has been the default choice for uncovering the ground truth. Suppose we wish to determine the true answer to the question: ``Is Philadelphia the capital of Pennsylvania?'' Condorcet's Jury Theorem suggests that if we elicit votes from a large crowd, the majority answer will be correct with high probability even if, on average, the crowd is only slightly more accurate than a random selection. However, in some domains the crowd can be highly inaccurate and experts may be in minority. For example, when the very question listed above is posed to real crowds, the majority answer is often (the incorrect) `yes'~\cite{DB17}. 

To circumvent this difficulty and uncover the ground truth even when the majority is wrong, \citet{PSM17} introduce the \emph{surprisingly popular} (SP) algorithm. This algorithm asks each individual not only what she thinks the answer is (the \emph{vote}), but also what fraction of the other participants she thinks will say yes/no (the \emph{prediction}). Then, instead of simply selecting the majority (i.e. popular) answer, the algorithm selects the answer that is \emph{surprisingly popular}, i.e., whose actual frequency in the votes is greater than its average predicted frequency. They show that as the crowd gets larger in the limit, this approach will provably recover the correct answer with probability $1$, even if the crowd is less accurate than a random selection on average. 

The intuition behind their algorithm, borrowed from their work, is as follows. Suppose there are two hypothetical worlds, one where Philadelphia is the capital and one where it is not. In the former world, a greater fraction (say $90\%$) would say `yes' than the fraction (say $60\%$) that would say `yes' in the latter. However, the $60\%$ of the people who believe the correct world is the former would predict the frequency of `yes' to be $90\%$, whereas the remaining $40\%$ would predict it to be $60\%$. This would make the average predicted frequency of `yes' to be somewhere between $60\%$ and $90\%$, higher than its actual frequency of $60\%$. In other words, the majority but incorrect answer `yes' would be surprisingly \emph{unpopular} while `no' would be surprisingly popular and correct. 

Several works have demonstrated the effectiveness of this approach in a wide range of domains~\cite{PSM17,LDV18,WLC19,PS19,RRCM20,MRP20}. Prediction questions have also been used to boost the accuracy of surveys on social networks~\cite{GBDK+18}. \citet{PSM17} show how to apply their approach to questions with non-binary votes and non-binary ground truth. When the true answer lurks among $r$ options, their approach requires each individual to predict the exact frequency of each of $r$ options among other individuals' votes. We are interested in ranked voting, i.e., when the ground truth is a ranking of $m$ alternatives. Note that in this case, the approach of \citet{PSM17}, which we refer to as \emph{surprisingly popular (SP) voting}, would require eliciting predictions in the form of a distribution over $r=m!$ options, which is clearly infeasible for even moderate values of $m$. Thus, the main research questions we address are:
\begin{quote}
    \emph{How do we extend surprisingly popular voting to effectively recover a ground truth ranking of alternatives? If we elicit partial vote and prediction, how do we aggregate them and what information-accuracy tradeoff does this offer?}
\end{quote}

\smallskip
\paragraph{Our contributions.} We focus on eliciting only \textit{ordinal} vote and prediction information. For the \emph{vote}, we ask individuals to provide their opinion of either just the top alternative of the ground truth ranking (\emph{Top}) or the full ground truth ranking (\emph{Rank}). For the \emph{prediction}, informally, we ask individuals to predict either just a single alternative (\emph{Top}) or a ranking of alternatives (\emph{Rank}) based on the other individuals' votes. The exact prediction elicited under various conditions is described in Section~\ref{sec:elicitation-formats}. In addition to these four elicitation formats, we use as benchmark two classical elicitation formats in which Top and Rank votes are elicited but no prediction is elicited. 
Because the SP algorithm of \citet{PSM17} does not work on partial votes and predictions, we first design a novel aggregation method for such partial information. 

Next, we conduct an empirical study with $720$ participants from Amazon's Mechanical Turk platform. We ask the participants questions on geography, movies, and artwork which admit a ground truth ranking of four alternatives and elicit their responses in the aforementioned six elicitation formats. We compare the different elicitation formats using four metrics: difficulty (measured through response time as well as perceived difficulty), expressiveness, error in recovering the ground truth top alternative, and error in recovering the ground truth ranking. 

Our results show that even when the vote and prediction information are individually no better than random guesses, by combining the two pieces of information SP voting performs significantly better. Further, it outperforms a whole slew of conventional voting rules which ignore prediction information and only aggregate the votes. We also observe that when it is necessary to choose between eliciting more complex vote information and eliciting more complex prediction information, the latter may be the right choice. 

\subsection{Related Work}\label{sec:related-work}

Our work builds on the SP voting approach of \citet{PSM17}. This approach in turn builds on its precursor, the Bayesian truth serum (BTS)~\cite{Prelec04}, which also uses participants' predictions, but for a different objective: to decide payoffs to the participants which incentivize them to honestly report their votes and predictions. 

Prediction markets~\cite{AFGH+08,CP10}, quadratic voting~\cite{LW18}, and peer prediction~\cite{MRZ05} are alternative approaches to recovering the ground truth, which, like SP voting, allow a minority of experts to override the majority opinion. Instead of eliciting participants' predictions of other participants' votes, prediction markets and quadratic voting ask participants to place a bet on their vote while peer prediction methods require them to participate in multiple tasks. 

These recent approaches stand in contrast to a large body of work on epistemic social choice~\cite{Piv19} and noisy voting~\cite{CPS16}, which build on the seminal work of \citet{Con85}, \citet{Gal07}, and \citet{Young88}. Some of this literature focuses on statistical models of errors in participants' votes such as the Mallows model, the Bradley-Terry model, the Thurstone-Mosteller model, and the Plackett-Luce model. However, all these models assume that a participant is ever-so-slightly more likely to report the correct option than an incorrect option. Hence, approaches based on these models can fail to recover the ground truth when the majority of the crowd is misinformed.

Finally, our work is reminiscent of a recent flurry of work on the elicitation-distortion tradeoff in computational social choice~\cite{MPSW19,AAZ19,MSW20,Kem20,ABFV20}. In this line of work, there is no ground truth; instead, participants have subjective preferences and the goal is to identify the decision that maximizes the social welfare. Rather than directly eliciting participants' utility functions, various elicitation formats are used to elicit partial preferences to analyze the tradeoff between the amount of information elicited and the approximation to social welfare (called distortion). Our work replaces the distortion with its counterpart, that is, the accuracy of recovering an underlying ground truth. 

\section{Model}\label{sec:model}

Let $A$ be a set of $m$ alternatives and $\calL(A)$ be the set of rankings over $A$. For a ranking $\sigma \in \calL(A)$ and $x \in \set{1,\ldots,m}$, let $\sigma(x)$ be the alternative  in the $x^{\text{th}}$ highest position in $\sigma$. 

SP voting uses a Bayesian model; in the following, we present a special case of the model for ranked voting. There exists a ground truth ranking $\pi^* \in \calL(A)$ drawn from a \emph{prior} $\prior$. There are $n$ voters; each voter $i$ observes a noisy ranking $\sigma_i \in \calL(A)$ drawn from a \emph{signal distribution} $\signal(\cdot | \pi^*)$. The voters know both the prior $\prior$ and the signal distribution $\signal(\cdot | \pi^*)$; however, the principal is unaware of both. Following \citet{PSM17}, we assume that $\prior(\pi),\signal(\sigma | \pi) > 0$ for all rankings $\sigma,\pi \in \calL(A)$ to avoid degeneracy. 

Conventional voting would ask each voter $i$ to simply report her observed noisy ranking $\sigma_i$ and use a  voting rule such as the Kemeny rule or Borda count  to aggregate the reported  rankings. SP voting additionally asks each voter $i$ to make inferences about the reports of other voters. Given her observed noisy ranking $\sigma_i$ and the prior $\calP$, voter $i$ can compute a posterior distribution over the ground truth, given by 
\[ 
\posterior(\pi^* | \sigma_i) 
= \frac{\signal(\sigma_i | \pi^*) \cdot \prior(\pi^*)}{\sum_{\pi' \in \calL(A)} \signal(\sigma_i | \pi') \cdot \prior(\pi')}.
\]
In turn, the voter can also infer a distribution over the noisy ranking $\sigma_j$ observed by another voter $j$:
\[
\inference(\sigma_j | \sigma_i) = \textstyle\sum_{\pi^* \in \calL(A)} \signal(\sigma_j | \pi^*) \cdot \posterior(\pi^* | \sigma_i).
\]
SP voting asks each voter $i$ to report not only her observed noisy ranking $\sigma_i$ (the \emph{vote}), but also her inferred distribution $\inference(\cdot | \sigma_i)$ over other voters' noisy rankings (the \emph{prediction}). Given these reports, for a ranking $\pi \in \calL(A)$, let $f(\pi) = \sum_{i=1}^n \one[\sigma_i = \pi]$ denote the number of voters who vote $\pi$ and $g(\cdot | \pi)$ denote the average of reported predictions $\inference(\cdot | \sigma_i)$ across all voters $i$ with $\sigma_i = \pi$. Then, the SP algorithm of \citet{PSM17} computes the prediction-normalized vote count for each possible ground truth $\pi$ as 
\begin{equation}\label{eqn:spvoting-rank}
\bar{V}(\pi) = f(\pi) \cdot {\textstyle\sum_{\pi' \in \calL(A)}} \frac{g(\pi' | \pi)}{g(\pi | \pi')}.
\end{equation}
The following result due to \citet{PSM17}, rephrased in our context, guarantees that the ground truth ranking will have the highest prediction-normalized vote count under the assumption that the highest posterior probability for ground truth ranking $\pi$ will be assigned by a voter who observes noisy ranking $\pi$. 
\begin{theorem}[\cite{PSM17}]
Suppose the prior $\calP$ and the signal distribution $\signal$ are such that $\posterior(\pi | \pi) > \posterior(\pi | \pi')$ for all distinct rankings $\pi,\pi' \in \calL(A)$. Then, we have that $\Pr[\pi^* \in \argmax_{\pi \in \calL(A)} \bar{V}(\pi)] \to 1$ as $n \to \infty$. 
\end{theorem}

\section{Elicitation Formats \& Aggregation Rules}\label{sec:elicitation-formats}

Note that the prediction requested from voter $i$, $\inference(\cdot | \sigma_i)$, is a distribution over $m!$ rankings. Eliciting this would undoubtedly place significant cognitive burden on the voter. Thus, our goal is to elicit partial vote and prediction information from the voters. 
Since eliciting numerical information is known to be difficult~\cite{Cam11}, we focus on eliciting ordinal information for prediction. We develop aggregation rules for recovering the ground truth from ordinal information and empirically evaluate the effectiveness of SP voting. 

\paragraph{Elicitation formats:} We focus on two types of vote reports, and for each of them, two types of prediction reports. Below we provide formal explanations of these formats in the context of our model. 
%
In the next section, we provide example phrasings that were used to pose the various questions to the participants in our empirical study. Let $r_i$ and $q_i$ respectively denote the vote and prediction reports submitted by voter $i$. 
\begin{itemize}
    \item \emph{Top vote:} Voter $i$ reports the top alternative in her observed noisy ranking, i.e., $r_i = \sigma_i(1)$.
    \begin{itemize}
        \item \emph{Top prediction:} Voter $i$ estimates the most frequent alternative among the other votes, i.e. $q_i = \argmax_{a \in A} \sum_{\sigma \in \calL(A) : \sigma(1) = a} \Pr_o(\sigma | \sigma_i)$. 
        \item \emph{Rank prediction:} Voter $i$ estimates the ranking of the alternatives by their frequency among the other votes, i.e., $q_i \in \calL(A)$ such that $\sum_{\sigma \in \calL(A) : \sigma(1) = q_i(x)} \Pr_o(\sigma | \sigma_i) \ge \sum_{\sigma \in \calL(A) : \sigma(1) = q_i(y)} \Pr_o(\sigma | \sigma_i)$ for all $x > y$.
    \end{itemize}
    \item \emph{Rank vote:} Voter $i$ reports her entire observed noisy ranking, i.e., $r_i = \sigma_i$.
    \begin{itemize}
        \item \emph{Top prediction:} Voter $i$ estimates the alternative that appears most frequently in the top position of the other votes. Formally, this is equivalent to the top prediction in case of a top vote: $q_i = \argmax_{a \in A} \sum_{\sigma \in \calL(A) : \sigma(1) = a} \Pr_o(\sigma | \sigma_i)$. 
        \item \emph{Rank prediction:} Voter $i$ estimates the most frequent ranking among the other votes, i.e., $q_i \in \argmax_{\sigma \in \calL(A)} \Pr_o(\sigma | \sigma_i)$. Note that this is different from the rank prediction in case of a top vote. 
    \end{itemize}
\end{itemize}

This gives rise to four elicitation formats, which we refer to as Top-Top, Top-Rank, Rank-Top, and Rank-Rank with the first component denoting the vote format and the second denoting the prediction format. As a benchmark, we use  Top-None and Rank-None, where top and rank votes are elicited, respectively, but no prediction information is elicited. 

\paragraph{Aggregation rules:} There are two difficulties in applying the SP algorithm of \citet{PSM17} --- maximizing $\bar{V}(\pi)$ given in \Cref{eqn:spvoting-rank} --- in our setting. 

First, the effectiveness of the approach depends on how accurately functions $f$ and $g$ from \Cref{eqn:spvoting-rank} match their expected values, which in turn depends on how large the number of voters is compared to the number of options among which the ground truth lurks. In our case, since the ground truth is one of $m!$ rankings, the approach would be ineffective unless each question is answered by a number of voters much larger than $m!$. Instead, we determine the ground truth comparison of each of $\binom{m}{2}$ pairs of alternatives independently by applying the algorithm from \Cref{eqn:spvoting-rank} on the relevant pairwise comparison data extracted from the reports of the voters. 

Second, even for comparing a pair of alternatives, \Cref{eqn:spvoting-rank} requires cardinal prediction information whereas our input is ordinal. We propose a simple parametric model in which, for each elicitation format, we use two parameters, $\alpha \in (0.5,1)$ and $\beta \in (0,0.5)$, to convert ordinal pairwise predictions into cardinal pairwise predictions to be utilized by the SP algorithm. In \Cref{sec:experimental-design}, we describe how we train these parameter values. The formal algorithm and its detailed description are provided in the appendix. 

Note that applying our algorithm for comparing each pair of alternatives independently results in a tournament, which we use for two prediction tasks: predicting the top alternative in the ground truth ranking and predicting the entire ground truth ranking. For the former task, we select the alternative that defeats the maximum number of other alternatives in the resulting tournament, breaking ties uniformly at random, and consider the frequency of predicting the correct top alternative. For the latter task, we compute the Kendall Tau distance of the tournament from the ground truth ranking. 

Finally, note that there are no prediction reports for Top-None and Rank-None and we consider a natural extension of SP voting. In particular, for Top-None, SP voting returns an acyclic tournament comparing alternatives by their plurality scores, and for Rank-None, it returns the (potentially cyclic) majority preference tournament. We then select an alternative/ranking as described earlier.


\section{Experiment Design} \label{sec:experimental-design}

To test the effectiveness of SP voting for recovering ranked ground truth with only ordinal elicitation, we conducted an empirical study by recruiting $720$ participants (turkers) from Amazon Mechanical Turk (MTurk), a popular crowdsourcing marketplace. An average turker spent about 15 minutes to complete the survey. The survey was designed as follows. 

\paragraph{Datasets.} To generate questions with an underlying ground truth comparison of alternatives, we used three datasets from three distinct \emph{domains}:
\begin{enumerate}
    \item The \emph{geography} dataset\footnote{Retrieved from \url{worldpopulationreview.com}} contains 230 countries with their 2019 population estimates according to the United Nations.
    \item The \emph{movies} dataset\footnote{Retrived from \url{boxofficemojo.com/chart/top_lifetime_gross}} contains 15,743 movies with their lifetime box-office gross earnings.
    \item The \emph{paintings} dataset\footnote{Generously provided by the authors of \citet{PSM17}.} contains 80 paintings with their latest auction prices.
\end{enumerate}

\paragraph{Questions.} In each domain, the numerical values associated with the alternatives allow a ground truth comparison among the alternatives. For each domain, we considered the top $50$ alternatives with the highest values. From these, we generated $20$ questions, each comparing four alternatives selected such that two  consecutive alternatives in the ground truth ranking were exactly $6$ ranks apart in the global ranking of all $50$ alternatives. Collectively, we had $60$ questions across all three domains. For each of the $60$ questions and  each of the $6$ elicitation formats described in \Cref{sec:elicitation-formats}, we elicited $20$ responses, generating a total of $7,200$ responses.

\begin{figure}[h!]
    \centering
    \scalebox{0.62}{
    \begin{tikzpicture}
      [
        start chain=p going right,
        every on chain/.append style={etape},
        every join/.append style={line},
        node distance= 0.75 and 0.75,
      ]
      {
        \node [on chain, join] {\includegraphics[width=0.5\textwidth]{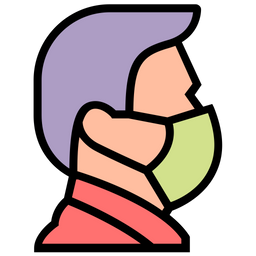}};
        \node [on chain, join] {Preview \& Consent};
        \node [on chain, join] {Tutorial\\(Elicitation Format 1)};
        \node [on chain, join] {5 Questions\\(Elicitation Format 1)};
        \node [on chain, join] {Review\\(Elicitation Format 1)};
        {[start branch=d going below]
    	  \node [on chain, join] {Tutorial\\(Elicitation Format 2)};
        }
        {[continue branch=d going left]
    	  \node [on chain, join] {5 Questions\\(Elicitation Format 2)};
    	  \node [on chain, join] {Review\\(Elicitation Format 2)};
    	  \node [on chain, join] {Quiz};
    	  \node [on chain, join] {Submit};
        }
      }
    \end{tikzpicture}}
    \caption{The workflow of a turker.}
    \label{fig:experiment}
\end{figure}
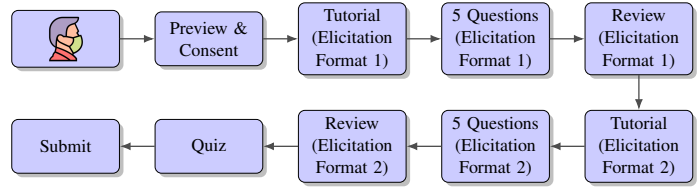

\paragraph{Turker assignment.} \Cref{fig:experiment} shows the workflow faced by a turker. Each of the $720$ turkers responded to $10$ questions split evenly among two randomly assigned elicitation formats. , 
The turkers were divided roughly equally between the $30$ ordered pairs of elicitation formats called \emph{treatments}. 
Further, as mentioned above, each question under each elicitation format was assigned to the same number of turkers. 

\paragraph{Tutorials.} As shown in \Cref{fig:experiment}, each set of five questions in a fixed elicitation format was preceded by a tutorial. The tutorial was designed specifically for the elicitation format and tested turkers' understanding of the vote and prediction formats. It contained a sample question along with pre-specified beliefs over the correct answer as well as over the other responses. Turkers had to successfully pass the tutorial by converting the given beliefs into the requested vote and prediction format in order to proceed to the questions. 

\paragraph{Reviews.} Each set of five questions was also succeeded by a review, which asked the turkers to rate the \emph{difficulty} (from Very Easy to Very Difficult) and \textit{expressiveness} (Very Little to Very Significant) of the elicitation format of the preceding questions. While we controlled the difficulty level of various questions from a given domain, as we show in \Cref{sec:results} the three domains themselves differed significantly in their difficulty. In anticipation of this and to ensure that the turkers' implicit comparison between their two assigned elicitation formats is not influenced by the domains, the study was designed such that the sequence of domains encountered by a turker in the first five questions precisely matched that in the next five questions. 

See the appendix for details such as the consent form, the tutorial for each domain, the review, and other details.

\begin{figure*}
\centering
\begin{minipage}[!b]{.32\linewidth}
    \includegraphics[width=\linewidth]{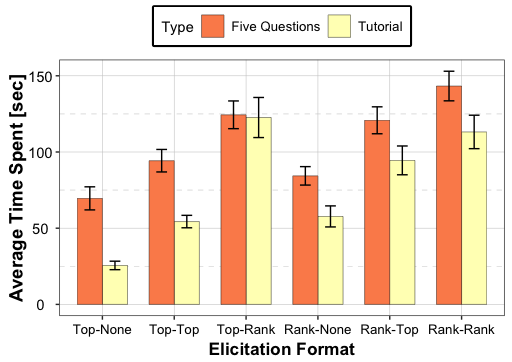}
    \caption{Average time spent.}
    \label{fig:meantime}
\end{minipage}\hspace{0.01\linewidth}%
\begin{minipage}[!b]{.32\linewidth}
    \includegraphics[width=\linewidth]{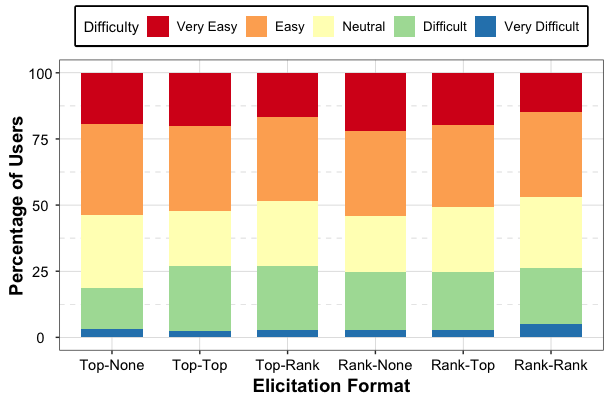}
    \caption{Perceived difficulty.}
    \label{fig:difficulty}
\end{minipage}\hspace{0.01\linewidth}%
\begin{minipage}[!b]{.32\linewidth}
    \includegraphics[width=\linewidth]{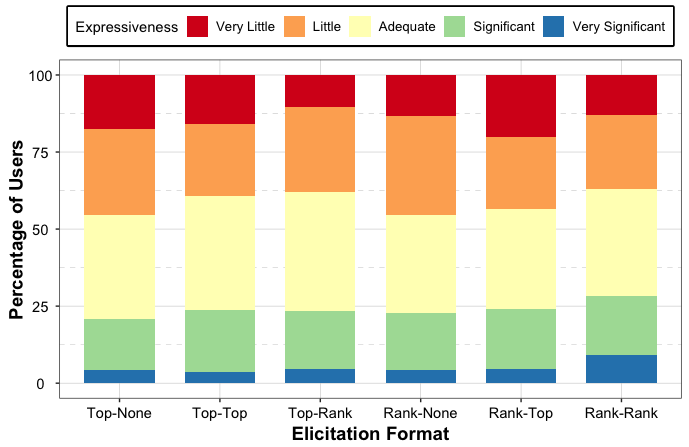}
    \caption{Perceived expressiveness.}
    \label{fig:expressiveness}
\end{minipage}
\end{figure*}

\paragraph{Response qualifications \& payment.} To ensure high-quality responses, in addition to providing training in the form of tutorials, we restricted participation in our study to turkers who had (a) at least $90\%$ approval rate on previous tasks, (b) at least $100$ completed tasks, and (c) the region set to US East (us-east-1) on MTurk. 

Finally, at the end of the survey, the turkers were required to answer a quiz, which repeated the four alternatives from the last question they answered and asked them to identify the alternative they chose or ranked first in their vote. The turkers were incentivized to answer the quiz correctly (see below). In our case, over $82\%$ of turkers passed the quiz.

The payment was divided into two parts. A \emph{base} payment of {50\textcent} was provided conditioned on completing the entire survey including all tutorials, questions, and reviews. A \emph{bonus} payment of {50\textcent} was provided conditioned on correctly answering the quiz question. 

\paragraph{Elicitation formats.} In \Cref{sec:elicitation-formats}, we discussed six elicitation formats and described what vote and prediction a given voter $i$ should submit as a function of her observed noisy ranking $\sigma_i$, the prior $\prior$, and the signal distribution $\signal$. In our empirical study, we design natural and intuitive phrasing to elicit the corresponding responses from the turkers. 

For example, consider a question which asks to compare four countries (United Kingdom, Vietnam, Russia, and Kenya) by their population. Under the Top-Rank elicitation format, the vote and prediction questions would be as follows:
\begin{itemize}
    \item \textbf{Part A (vote):} \emph{Which country do you think is the most populated among the following?}
    \item \textbf{Part B (prediction):} \emph{Imagine that other participants will also answer Part A. How do you think the following countries will be ordered from the most common response (top) to the least common (bottom)?}
\end{itemize}

See the appendix for sample phrasings for all six elicitation formats and screenshots from our user interface. 

\paragraph{Training.} Recall that in our aggregation method, for each elicitation format, we use two parameters, $\alpha \in (0.5,1)$ and $\beta \in (0,0.5)$, to convert ordinal predictions into cardinal predictions that can be then used in the SP algorithm. To learn effective values of these parameters, we split the dataset into a training and a test set. For each elicitation format, we selected $5$ questions from each of three domains, reserving the remaining $15$ questions from each domain for the test set. Using these $15$ questions, we performed a grid search over $\alpha$ ranging from $0.55$ to $0.95$ in increments of $0.025$ and $\beta$ ranging from $0.05$ to $0.45$ in increments of $0.025$ and selected the values with the lowest mean squared error.




\section{Results}\label{sec:results}
In this section, we present our results averaged across all three domains. In the appendix, we present more detailed results averaged across each domain separately. All confidence intervals shown are $95\%$ intervals. We compare the elicitation formats using four key metrics: difficulty (i.e. cognitive burden), expressiveness, error in predicting the ground truth top alternative, and error in predicting the ground truth ranking. 

\subsection{Difficulty \& Expressiveness}\label{sec:burden}
We measure the following three metrics.
\begin{itemize}
    \item \emph{Response time:} Response time is known to be a good objective proxy for the cognitive load associated with a task~\cite{Rau92}. We measure the amount of time spent by the turkers on the tutorials and questions of the elicitation format. 
    \item \emph{Perceived difficulty:} As a subjective indicator of difficulty, we consider the perceived difficulty reported by the turkers (from Very Easy to Very Difficult) during the review stage of the elicitation format. 
    \item \emph{Perceived expressiveness:} Expressiveness indicates the amount of information that the turkers felt they were able to convey through the elicitation format (from Very Significant to Very Little).  
\end{itemize}

\Cref{fig:meantime} shows the average time spent by the workers on the tutorial and on an average question under the six elicitation formats along with $95\%$ confidence intervals (lower is better). We observe a statistically significant trend: when we fix a vote format (say Top or Rank), the average time spent increases for both tutorials and questions as we make the prediction format more complex (None $\to$ Top $\to$ Rank). 
In the appendix, we show the average time spent for each domain and observe that the choice of the domain does not significantly affect it regardless of the elicitation format. 

\Cref{fig:difficulty} and \Cref{fig:expressiveness} respectively show the reported distributions of perceived difficulty (easier is better) and perceived expressiveness (higher is better). Interestingly, the turkers found the six elicitation formats to be of very similar difficulty and similar expressiveness.

\begin{figure*}[t]
\centering
\begin{minipage}[!b]{.48\linewidth}
    \centering
    \includegraphics[width=0.8\linewidth]{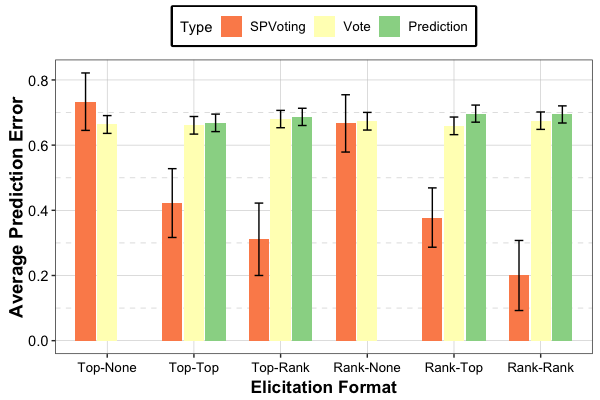}
    \caption{Average error in predicting the top alternative in the ground truth. By combining both the vote and predictions, SP voting achieves a much lower error than in either piece of information.}
    \label{fig:top-all-sip}
\end{minipage}\hspace{0.03\linewidth}%
\begin{minipage}[!b]{.48\linewidth}
    \centering
    \includegraphics[width=0.8\linewidth]{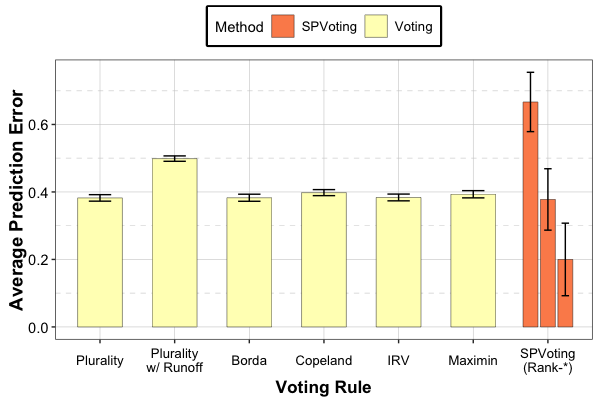}
    \caption{Comparing SP voting with conventional voting for predicting the top alternative. Incorporating the prediction reports helps SP voting significantly outperform conventional voting.}
    \label{fig:top-all-voting}
\end{minipage}\\%
\begin{minipage}[!b]{.48\linewidth}
    \centering
    \includegraphics[width=0.8\linewidth]{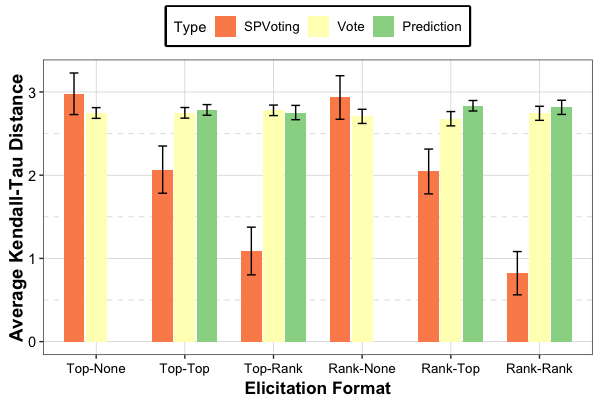}
    \caption{Average error in predicting the ground truth ranking. By combining both the vote and prediction information, SP voting achieves a much lower error than in either piece of information.}
    \label{fig:kt-all-sip}
\end{minipage}\hspace{0.03\linewidth}%
\begin{minipage}[!b]{.48\linewidth}
    \centering
    \includegraphics[width=0.8\linewidth]{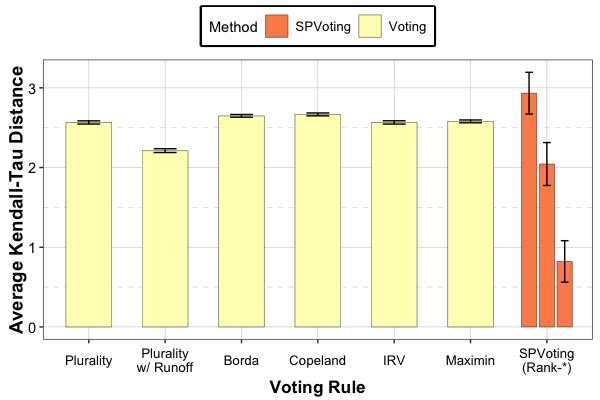}
    \caption{Comparing SP voting with conventional voting for predicting the ground truth ranking. Incorporating the prediction reports helps SP voting significantly outperform conventional voting.}
    \label{fig:kt-all-voting}
\end{minipage}
\end{figure*}

\subsection{Predicting the Ground Truth Top Alternative}\label{sec:predicting-top}

We now turn to analyzing how effectively the different elicitation formats help us predict the ground truth. In addition to measuring the error of the ground truth estimate returned by our algorithm, we also measure the error in the input votes and predictions themselves. Note that every vote and prediction is an estimate of some truth (either the ground truth or a summary statistic of the other votes); thus, its error can be measured with respect to the truth it is attempting to uncover. 

First, we consider predicting simply the top alternative in the ground truth ranking. For our algorithm as well as for the input votes and predictions, we use, as error measure, the frequency of incorrectly guessing the top alternative of the truth they attempt to estimate. \Cref{fig:top-all-sip} shows the average prediction errors for various elicitation formats (lower is better).\footnote{SP voting errors are obtained by averaging over $60$ elections associated with $60$ questions. Vote/Prediction errors are averaged over $1200$ responses and have narrower confidence intervals.} We remind the reader that the effectiveness of SP voting should be judged based only on elicitation formats which include some prediction information.

Given four alternatives, selecting an alternative uniformly at random would result in a prediction error of $0.75$. Interestingly, both the vote and prediction reports individually have average error around this benchmark.
Yet, by combining these two pieces of individually erroneous information, SP voting is able to achieve significantly lower error. This is not surprising because SP voting approach is design precisely to pick out the minority of experts lurking among a majority of non-experts by combining vote and prediction information. Moreover, for a fixed type of vote (either Top or Rank), as the prediction formats become more complex (None $\rightarrow$ Top $\rightarrow$ Rank), the performance of SP voting improves.

\Cref{fig:top-all-voting} compares SP voting to several standard voting rules including Plurality, Plurality with Runoff, Borda, Copeland, Instant Runoff Voting (IRV), and Maximin Rule, which ignore the prediction information and simply aggregate the vote information in a democratic manner.\footnote{See \cite{brandt2016handbook} for definitions of these rules.} 
The conventional voting rules run on elections containing votes from three elicitation formats (Rank-None, Rank-Top, and Rank-Rank) whereas SP voting runs on each elicitation format individually. We can see that for Rank-Rank, SP voting (rightmost orange bar) outperforms all conventional voting rules, despite having access to just a third of the samples. This indicates that the prediction information helps significantly. 
%
%

These observations hold even when we consider each domain separately. These results are provided in the appendix. 

\subsection{Predicting the Ground Truth Ranking}\label{sec:predicting-rank}

We now consider predicting the full ground truth ranking. For SP voting result as well as the individual votes and predictions, we use the Kendall-Tau (KT) distance to measure the error of the SP voting result, votes, and predictions compared to the true ranking they aim to estimate. 
\Cref{fig:kt-all-sip} shows the average KT distance for different elicitation formats (lower is better). Given four alternatives, selecting a uniformly random ranking will have an average KT distance of $3$. Both the votes and prediction reports have average error around this benchmark.
Similar to predicting the top alternative, SP voting produces significantly lower average error by combining these two noisy pieces of information. Moreover, for each vote format (either Top or Rank), as the prediction report becomes more expressive (None $\to$ Top $\to$ Rank) the average error of SP voting decreases.

Finally, we compare SP voting with standard voting rules (\Cref{fig:kt-all-voting}) in terms of the average KT distance and find that SP voting again outperforms all voting rules for Rank-Rank.

\subsection{Prediction vs. Vote}

Our results illustrate the importance of prediction in recovering the ground truth. While eliciting ranked votes and predictions (Rank-Rank) achieves the lowest error, an intriguing question arises when we seek to choose an elicitation format that provides a reasonable tradeoff between accuracy and difficulty/expressiveness.
\Cref{fig:top-all-sip,fig:kt-all-sip} show that Top-Rank significantly outperforms Rank-Top while both formats are comparable in terms of response time, perceived difficulty, and perceived expressiveness. Thus, if we wish to choose an elicitation format slightly more complex than Top-Top, making the prediction more expressive is more promising than that of the vote. The same observation holds when comparing Top-Top versus Rank-None. This shows that when a tradeoff between more complex vote and more complex prediction is necessary, eliciting more complex prediction may be better. 




\section{Discussion}\label{sec:discussion}
We extended surprisingly popular voting to recover a ground truth ranking of alternatives and, through a crowdsourcing study across different domains, showed that it outperforms conventional voting approaches without significantly increasing elicitation. 
In our study, the ground truth is a ranking over four alternatives, and a challenging future direction is to extend this approach to rankings with more than four alternatives. For a large number of alternatives, any practical elicitation scheme would ask the voters to report a partial rank over the alternatives, which will make it challenging to design aggregation rules for such partial ranks.

Another interesting direction would be to derive theoretical performance guarantees for surprisingly popular voting when the number of participants is finite (the results of \citet{PSM17} hold only in the limit) and when only partial votes and predictions are elicited (this may require assuming a parametric signal distribution such as the Mallows model). 


\paragraph{Acknowledgements.}
 The authors were partly supported by NSF grant \#1850076 (HH), a postdoctoral fellowship from Columbia DSI (DM), and an NSERC Discovery Grant (NS).

\bibliographystyle{named}
\bibliography{references}

\appendix
\onecolumn
\section*{\centering Appendix}

\section{Additional Material for Section \ref{sec:experimental-design}}

\subsection{Experiment Design}
Since we are interested in recovering the rank of four alternatives, any question where the four alternatives are ranked very close to each other in the ranked list of $50$ alternatives, would be a hard question for most of the turkers. For this reason, we decided to formulate each question by selecting four alternatives where the distance between successive alternatives is exactly six. For example, some  questions had alternatives ranked $(1,7,13,19), (3,9,15,21),\ldots,(31,37,43,49)$. The consecutive alternatives were kept exactly $6$ distance apart to control the difficulty level of the questions. 
The first $16$ questions have alternatives ranked $(1,7,13,19), (3,9,15,21),\ldots,(31,37,43,49)$. The last four questions have alternatives ranked $(2,8,14,20), (12,18,24,30), (22,28,34,40), (32,38,44,50)$.  

\paragraph{Elicitation Formats.} Below, we provide the phrasings used in our study to elicit the various types of responses from a turker. As an example question, consider a question which asks to compare four countries (United Kingdom, Vietnam, Russia, and Kenya) by their population. 


\begin{enumerate}
    \item {Top-None}: A turker is provided with four choices and is asked to vote the best option according to her opinion. 
    \begin{itemize}
        \item \textbf{Part A (vote):} \emph{Which country do you think is the most populated among the following?}
    \end{itemize}
    The turker is just asked one question and there is no additional prediction question regarding others' opinions.
    \item {Top-Top}: A turker is asked two questions in this format. She is asked to vote her top choice as in  format {Top-None}. Moreover, she is asked a prediction question about the votes of other participants. 
    \begin{itemize}
        \item \textbf{Part A (vote):} \emph{Which country do you think is the most populated among the following?}
        \item \textbf{Part B (prediction):} \emph{Imagine that other participants will also answer Part A. Which of the following four countries do you think will be the most common response?}
    \end{itemize}
    \item {Top-Rank}: Like {Top-Top}, this rule also asks a turker two questions. However, the prediction question is different, and asks to rank the four choices based on the votes of other participants.
 \begin{itemize}
        \item \textbf{Part A (vote):} \emph{Which country do you think is the most populated among the following?}
        \item \textbf{Part B (prediction):} \emph{Imagine that other participants will also answer Part A. How do you think the following countries will be ordered from the most common response (top) to the least common (bottom)?}
    \end{itemize}
    \item {Rank-None}: This elicitation format asks the turker to order the four choices based on her own opinion. 
   \begin{itemize}
        \item \textbf{Part A (vote):} \emph{How do you think the following four countries should be ordered from the most-populated (top) to the least-populated (bottom)?}
    \end{itemize}
    \item {Rank-Top}: This format also asks a turker to rank four choices and same as in Rank-None. Additionally, it asks the turker a prediction question about the votes of other participants.
     \begin{itemize}
        \item \textbf{Part A (vote):} \emph{How do you think the following four countries should be ordered from the most-populated (top) to the least-populated (bottom)?}
        \item \textbf{Part B (prediction):} \emph{Imagine that other participants will also answer Part A. In your opinion, which country will be the most common top choice?}
    \end{itemize}
    \item {Rank-Rank}: In this format, both the vote and prediction questions ask the turker to rank the four choices.
     \begin{itemize}
        \item \textbf{Part A (vote):} \emph{How do you think the following four countries should be ordered from the most-populated (top) to the least-populated (bottom)?}
        \item \textbf{Part B (prediction):} \emph{Imagine that other participants will also answer Part A. In your opinion, which will be the most common ordering of the following countries?}
    \end{itemize}
    
\end{enumerate}


\section{Details of the Aggregation Method}\label{sec:method}

We now discuss our aggregation method that takes as input pairs of (vote, prediction) reports from the voters and returns either a rank or a single alternative. Recall that we considered six different elicitation formats with different types of votes and prediction reports. Both the vote and the prediction report can be either the top alternative or a ranking over the alternatives. Our aggregation method is a general aggregation rule and works when the votes and prediction reports are either ranked choices or top alternatives. 
%
At a high level, our method (algorithm \ref{algo:aggr}) considers pairs of alternatives (say $(a,b)$) in turn, extracts the relevant information for that pair from the input, and determines whether $a \succ b$ or not in the true ranking $\pi_\star$. 




Algorithm \ref{algo:extr} describes how to extract the relevant information about a pair $(a,b)$ from the input. 
We first  describe how to extract the relevant votes about the pair $(a,b)$ from the input. 
Consider a vote $r_i$ from voter $i$. If $r_i$ is a rank over the alternatives, then we set $r_i^{(a,b)}$ either $1$ or $0$ based on whether  $a \succ_{r_i} b$ or not. On the other hand, if $r_i$ is just an alternative, then set $r_i^{(a,b)}$ to either $1$ or $0$ depending on whether the reported alternative equals $a$ or $b$. If  the top alternative $r_i$ is neither $a$ nor $b$, we just discard report of voter $i$ for determining the order for the pair of alternatives $a$, and $b$. 

In order to extract the relevant prediction report about the pair $(a,b)$, note that $q_i$ can be either a rank or an alternative, and we want to convert it to a probability estimate of $P(a \succ b | a \succ b)$ or $P(a \succ b | b \succ a)$. So algorithm \ref{algo:extr} takes as input two additional parameters $\alpha$, and $\beta$ and they are used to determine the value of the probability estimates. 
If $q_i$ is a rank over the alternatives, then we first choose either $\alpha$ or $\beta$ depending on the value of $r^{(a,b)}_i$. Call this choice $p$. 
Then set $q_i^{(a,b)}$ either $p$ or $1-p$ based on whether  $a \succ_{q_i} b$ or not. On the other hand, if $q_i$ is just an alternative, then set $q_i^{(a,b)}$ to either $p$ or $1-p$ depending on whether the prediction equals $a$ or $b$. Finally, in case the predicted alternative is neither $a$ nor $b$, we set $q_i^{(a,b)}$ to $1/2$.

After extracting the relevant information about the pair $(a,b)$, algorithm \ref{algo:aggr} executes the SP algorithm on two possible comparisons between $a$ and $b$ -- $a \succ b$ and $b \succ a$. In particular, we compute the \emph{prediction-normalized vote} for $a \succ b$ and $b \succ a$, denoted as $\bar{V}(a \succ b)$ and $\bar{V}(b \succ a)$ respectively, and choose whichever has higher normalized vote. The formula for the prediction-normalized vote for a comparison (say $a \succ b$) is given as
$$
\bar{V}(a \succ b) = f(a \succ b) \sum_i  \frac{g\left(r_i^{(a,b)}|a \succ b \right)}  {g\left(a \succ b|r_i^{(a,b)}\right)}
$$
Here $f(a \succ b)$ is the true frequency of the order $a \succ b$ among the voters, which we approximate from the votes $\{r^{(a,b)}_i\}_{i \in [n]}$. Similarly, the conditional probabilities are approximated from the prediction reports $\{q^{(a,b)}_i\}_{i \in [n]}$. For more details about the guarantees of this algorithm, the reader is referred to the supplementary materials of \cite{PSM17}.


\begin{algorithm}
\DontPrintSemicolon
\KwInput{Information reports $\{r_i\}_{i \in [n]}$, prediction reports $\{q_i\}_{i \in [n]}$, and probabilities $\alpha > 0.5$ and $\beta < 0.5$.}
\For{each pair of alternatives $(a,b)$}{
$\left(\{r^{(a,b)}_i, q^{(a,b)}_i\}_{i \in [n]}\right) \leftarrow \textrm{Extract-Reports}(\{r_i,q_i\}_{i \in [n]}, (a,b), \alpha, \beta)$\\
 \tcc{Signal $1$ (resp. $0$) corresponds to $a \succ b$ (resp. $b \succ a$).}
 $N_{a \succ b} = \set{j : r_j^{(a,b)} = 1}$\\
 $N_{b \succ a} = \set{j : r_j^{(a,b)} = 0}$\\
 $f(a \succ b) = \sum_i \one\set{r_i^{(a,b)} = 1} / (\abs{N_{a \succ b}} + \abs{N_{b \succ a}})$\\
 $f(b \succ a) = 1 - p_{\star}(a \succ b)$\\
 $g(1|1) = \frac{1}{\abs{N_{a \succ b}}} \sum_{i \in N_{a \succ b}} q_i \textrm{  and  } P(0|1) = 1 - P(1|1)$\\
 $g(1|0) = \frac{1}{\abs{N_{b \succ a}}} \sum_{i \in N_{b \succ a}} q_i \textrm{  and  } P(0|0) = 1 - P(1|0)$\\
 \tcc{Compute prediction-normalized vote}
 \begin{flalign*}
 &\bar{V}(a \succ b) = f(a \succ b) \sum_i  \frac{g\left(r_i^{(a,b)}|1 \right)}  {g\left(1|r_i^{(a,b)}\right)}&&\\
 &\bar{V}(b \succ a) = f(b \succ a) \sum_i  \frac{g\left(r_i^{(a,b)}|0 \right)}  {g\left(0|r_i^{(a,b)}\right)}&&
 \end{flalign*}
}
$T \leftarrow \phi$ \tcc{Create a tournament}
\For{each pair of alternatives $(a,b)$}{
\tcc{Ties are broken u.a.r.}
\uIf{$\bar{V}(a \succ b) > \bar{V}(b \succ a)$} 
{
    $T \leftarrow T \cup a \succ b$\\
}
\Else{$T \leftarrow T \cup b \succ a$\\}
}
\Return{$T$}
\caption{SP Voting \label{algo:aggr}}
\end{algorithm}



\begin{algorithm}[!h]
\DontPrintSemicolon
\KwInput{Information reports $\{r_i\}_{i \in [n]}$, prediction reports $\{q_i\}_{i \in [n]}$, pair $(a,b)$, and probabilities $\alpha > 0.5$, and $\beta < 0.5$.}

    \For{$i = 1, \ldots, n$}
    {
        \tcc{Extract information report}
        \uIf{$r_i$ is a rank}
        {Set $r^{(a,b)}_i = \left\{\begin{array}{cc}
            1  &  \textrm{ if } a \succ_{r_i} b\\
            0  &  \textrm{ o.w. }
        \end{array} \right.$}
        \uElseIf{$r_i$ is a top alternative and $r_i \in \set{a,b}$}
        {Set $r^{(a,b)}_i = \left\{\begin{array}{cc}
            1  &  \textrm{ if } r_i = a\\
            0  &  \textrm{ if } r_i = b
        \end{array} \right.$
        }
        \Else{Ignore $(r_i,q_i)$ for determining $a \succ b$}
        
        \tcc{Extract prediction report}
        \uIf{$q_i$ is a rank}
        {Set $q^{(a,b)}_i = \left\{\begin{array}{cc}
            \alpha  &  \textrm{ if } a \succ_{q_i} b \textrm{ and } r_i^{(a,b)} = 1\\
            1-\alpha & \textrm{ if } b \succ_{q_i} a \textrm{ and } r_i^{(a,b)} = 1 \\
            1-\beta & \textrm{ if } a \succ_{q_i} b \textrm{ and } r_i^{(a,b)} = 0\\
            \beta &  \textrm{ o.w. }
        \end{array} \right.$}
        \uElseIf{$q_i$ is a top alternative and $q_i \in \set{a,b}$}{
        Set $q^{(a,b)}_i = \left\{\begin{array}{cc}
        \alpha  &  \textrm{ if } q_i = a \textrm{ and } r_i^{(a,b)} = 1\\
            1-\alpha & \textrm{ if } q_i = b \textrm{ and } r_i^{(a,b)} = 1 \\
            1-\beta & \textrm{ if } q_i = a  \textrm{ and } r_i^{(a,b)} = 0\\
            \beta & \textrm{ o.w. }
        \end{array} \right.$
        }
        \Else{Set $q^{(a,b)}_i = 1/2$.}
    }
\Return{$\left(\{r^{(a,b)}_i, q^{(a,b)}_i\}_{i \in [n]}\right)$}
\caption{Extract-Reports \label{algo:extr}}
\end{algorithm}

\newpage

\section{Missing Figures for Different Domains}
\subsection{Mean Time Spent}
\begin{figure}[!h]
    \centering
    \includegraphics[width=0.5\linewidth]{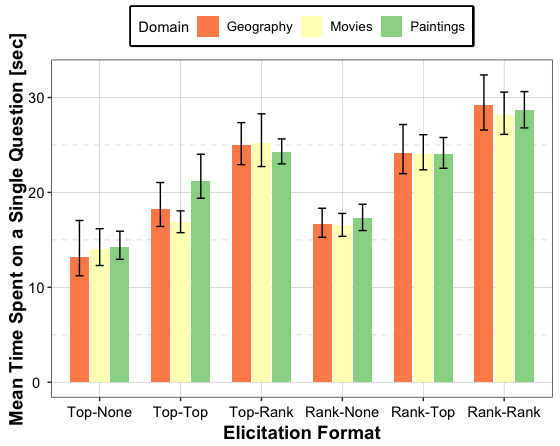}
    \caption{Average time-spent on a question, for different elicitation formats, grouped by domain.}
    \label{fig:meantime-domain}
\end{figure}

Figure \ref{fig:meantime-domain} shows the average time spent by a turker on a single question, for three different domains. For each domain, we observe the same pattern. For a fixed type of vote (either Top or Rank), as we ask more complex prediction reports ($\text{none} \rightarrow \text{top} \rightarrow \text{rank}$), the particular elicitation format requires more time to answer one question. 

\newpage

\subsection{Predicting Top Alternative for Different Domains}
\begin{figure*}[!h]
\begin{minipage}{0.95\linewidth}
    \centering
    \includegraphics[width=\linewidth]{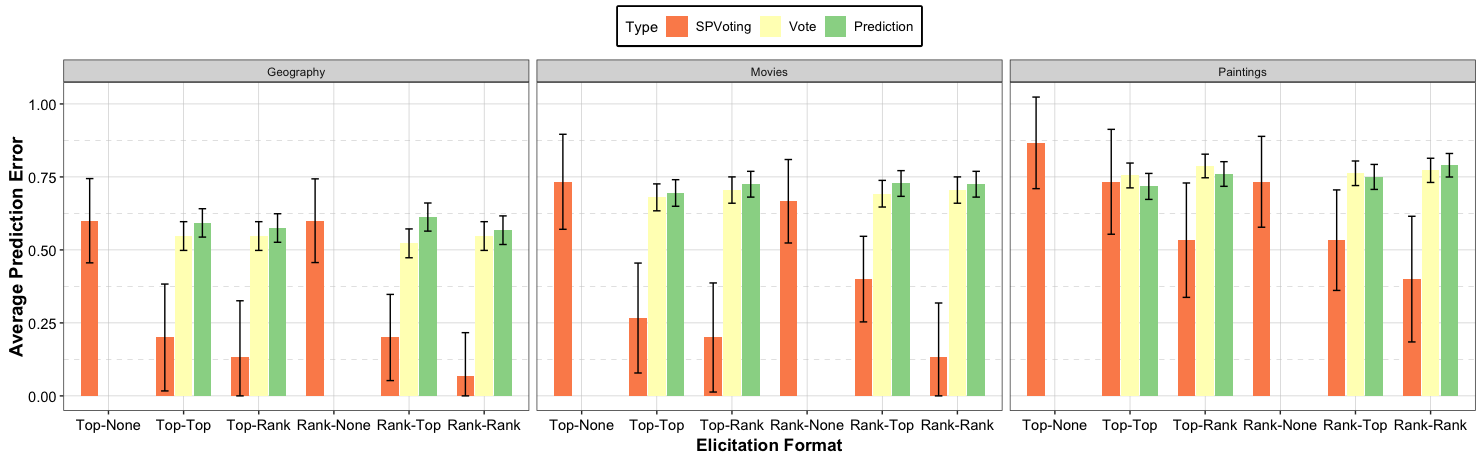}
    \caption{Average error in predicting the top alternative of the ground truth ranking, across different elicitation formats and  different domains.}
    \label{fig:avg-top-domains}
\end{minipage}
\begin{minipage}{0.95\linewidth}
    \centering
    \includegraphics[width=\linewidth]{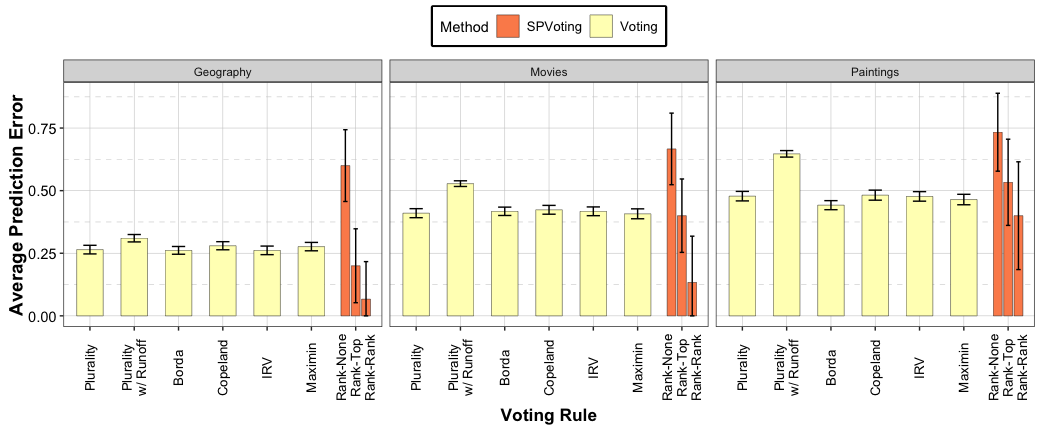}
    \caption{Average error in predicting the top alternative of the ground truth ranking, for different voting rules, and SP voting on three elicitation formats (Rank-None, Rank-Top, and Rank-Rank).}
    \label{fig:top-domains-voting}
\end{minipage}
\end{figure*}

Figure \ref{fig:avg-top-domains} show the average error in predicting the top alternative of the ground truth ranking for different elicitation formats and different domains. For each domain, we see that, for a fixed type of vote (Top or Rank) as we make the prediction reports more complex, the average prediction error generally goes down. In particular, except for the domain Paintings, the following orders always hold among the elicitation formats: $\text{Top-None} > \text{Top-Top}$ and $\text{Rank-None} > \text{Rank-Top}$. 

Figure \ref{fig:top-domains-voting} compares our method with six conventional voting rules in terms of predicting the top alternative of the ground truth. We see the same phenomenon as we saw when all questions were combined. SP voting trained on just Rank-Rank elicitation format, outperforms all six voting rules for the domains Geography and Movies.

\newpage
\subsection{Predicting  Ground Truth Ranking for Different Domains}
\begin{figure*}[!h]
\begin{minipage}{0.95\linewidth}
    \centering
    \includegraphics[width=\linewidth]{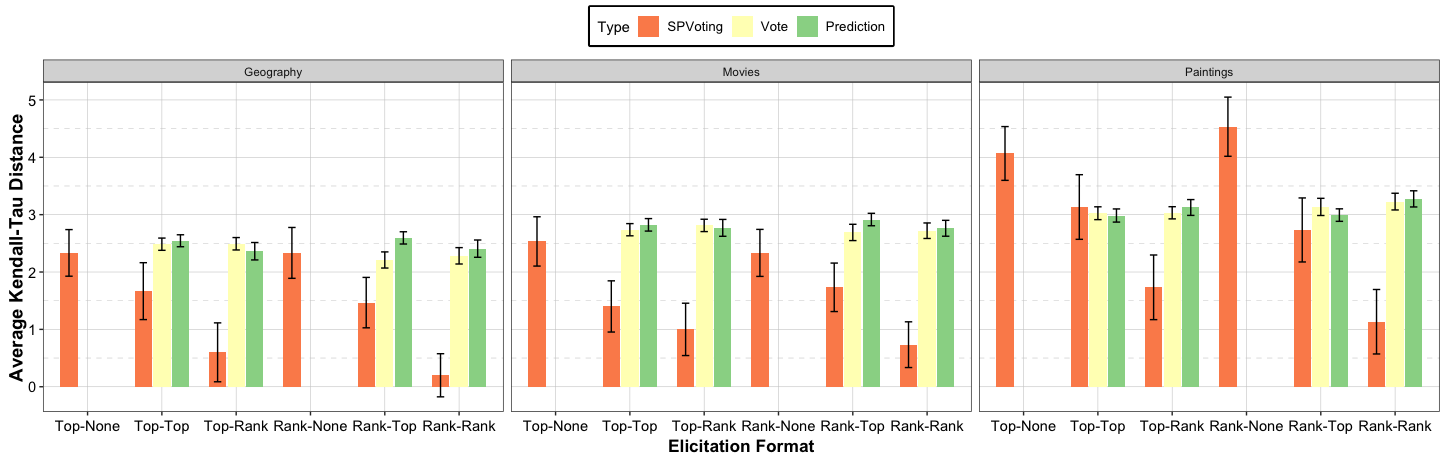}
    \caption{Average Kendall-Tau distance from the true rankings, across different elicitation formats, and different domains.}
    \label{fig:avg-kt-domains}
\end{minipage}
\begin{minipage}{0.95\linewidth}
    \centering
    \includegraphics[width=\linewidth]{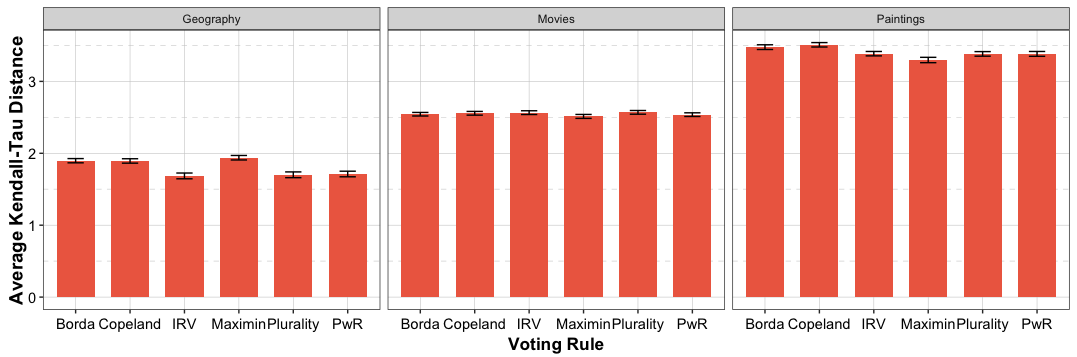}
    \caption{Average Kendall-Tau distance from the true rankings, for different voting rules, and SP voting on three elicitation formats (Rank-None, Rank-Top, and Rank-Rank).}
    \label{fig:kt-domains-voting}
\end{minipage}
\end{figure*}

Figure \ref{fig:avg-kt-domains} shows the average Kendall-Tau distance from the ground truth ranking for different elicitation formats and different domains. For each domain, we see that, for a fixed type of vote (Top or Rank) as we make the prediction reports more complex, the average prediction error generally goes down. In particular, except for the domain Paintings, the following orders always hold among the elicitation formats: $\text{Top-None} > \text{Top-Rank}$ and $\text{Rank-None} > \text{Rank-Rank}$. 

Figure \ref{fig:kt-domains-voting} compares our method with six conventional voting rules in terms of the average Kendall-Tau distance from the underlying true ranking. We see the same phenomenon as we saw when all questions were combined. SP voting trained on just Rank-Rank elicitation format, outperforms all six voting rules for all the domains.

\clearpage


\section{Screenshots from Our User Interface}
In this section, we provide screenshots of  different pages of our user interface.
\subsection{Preview and Consent Form}
\begin{figure*}[h!]
     \centering
     \begin{subfigure}[b]{0.45\textwidth}
         \centering
         \includegraphics[width=\textwidth]{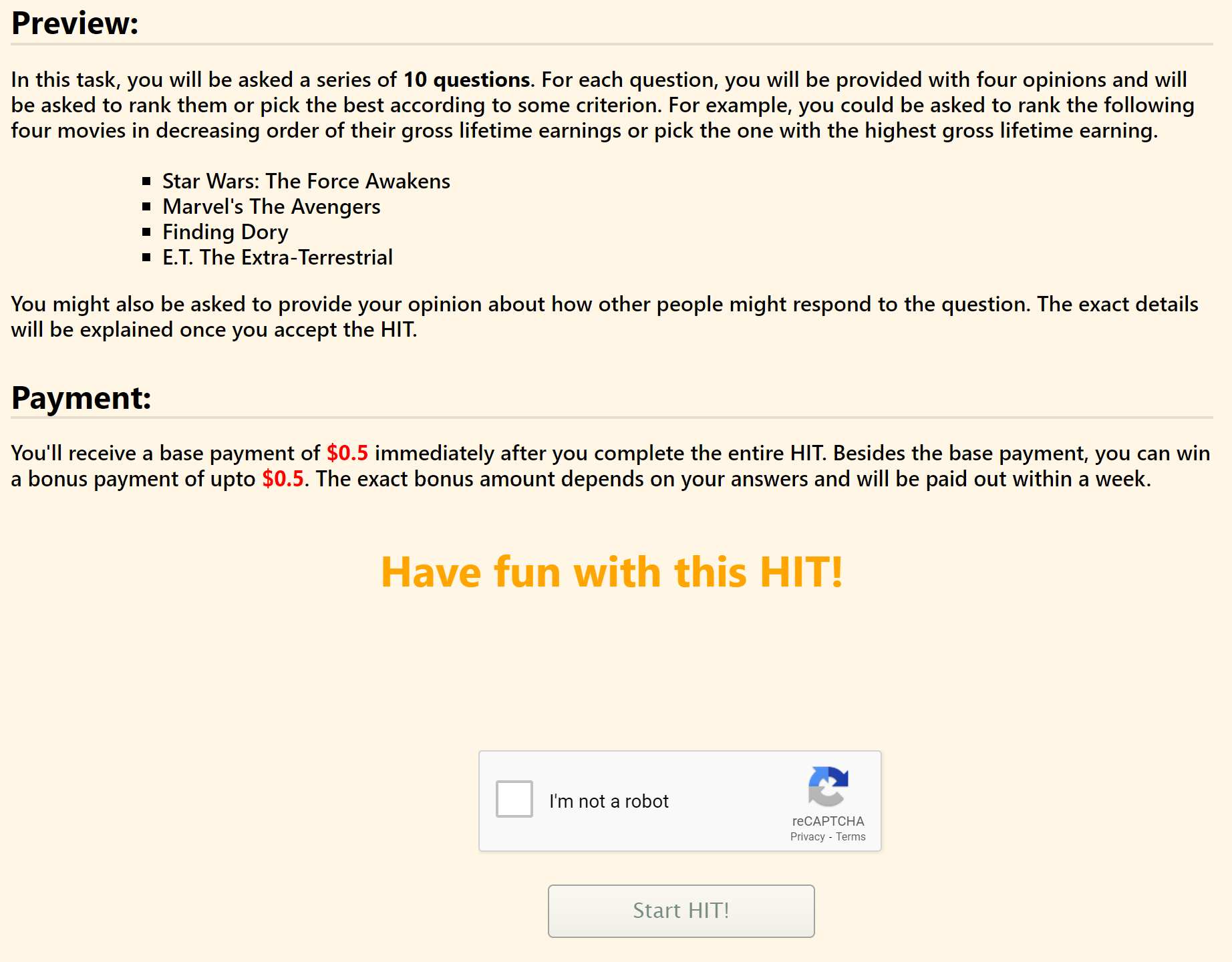}
     \end{subfigure}
     \hfill
     \begin{subfigure}[b]{0.45\textwidth}
         \centering
         \includegraphics[width=\textwidth]{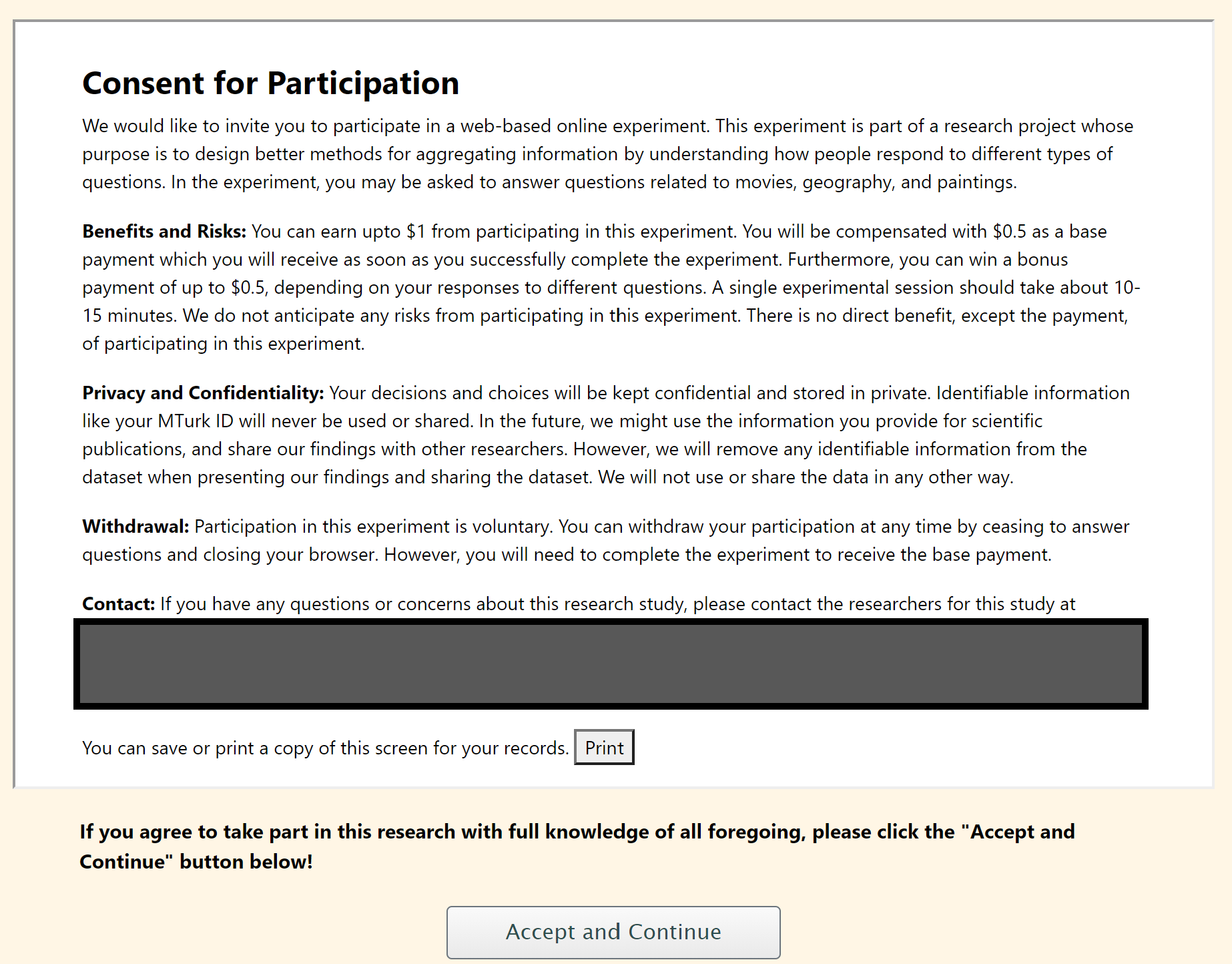}
     \end{subfigure}
        \caption{Preview and consent form \label{fig:preview-consent}}
\end{figure*}

Figure \ref{fig:preview-consent} shows the Preview and the Consent pages. After accepting our HIT, a turker first sees the Preview page. The turker needs to accept the consnt for participation before continuing with the tutorials.

\subsection{Tutorials}
\begin{figure*}[h!]
     \centering
     \begin{subfigure}[b]{0.45\textwidth}
         \centering
         \includegraphics[width=\textwidth]{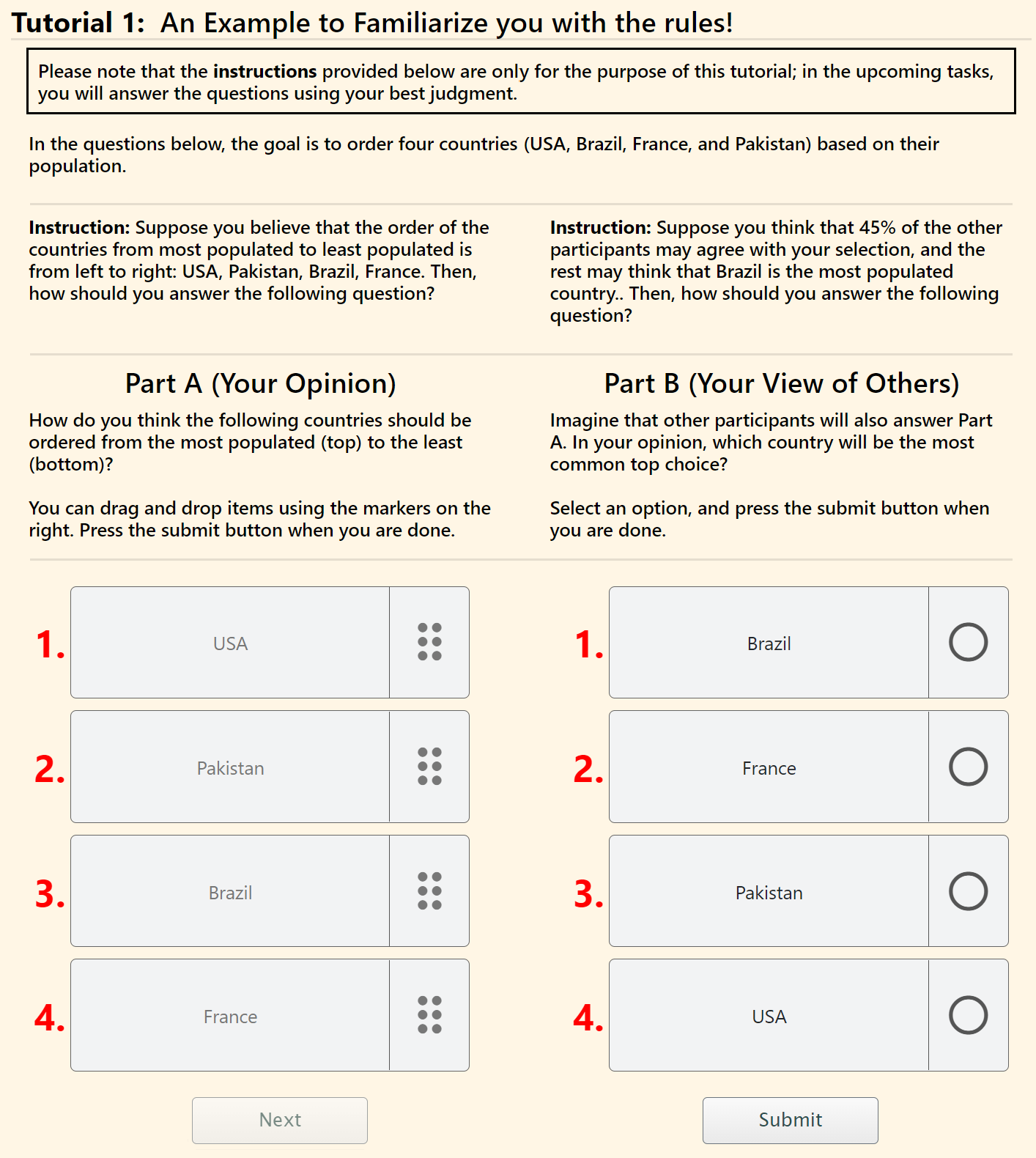}
     \end{subfigure}
     \hfill
     \begin{subfigure}[b]{0.45\textwidth}
         \centering
         \includegraphics[width=\textwidth]{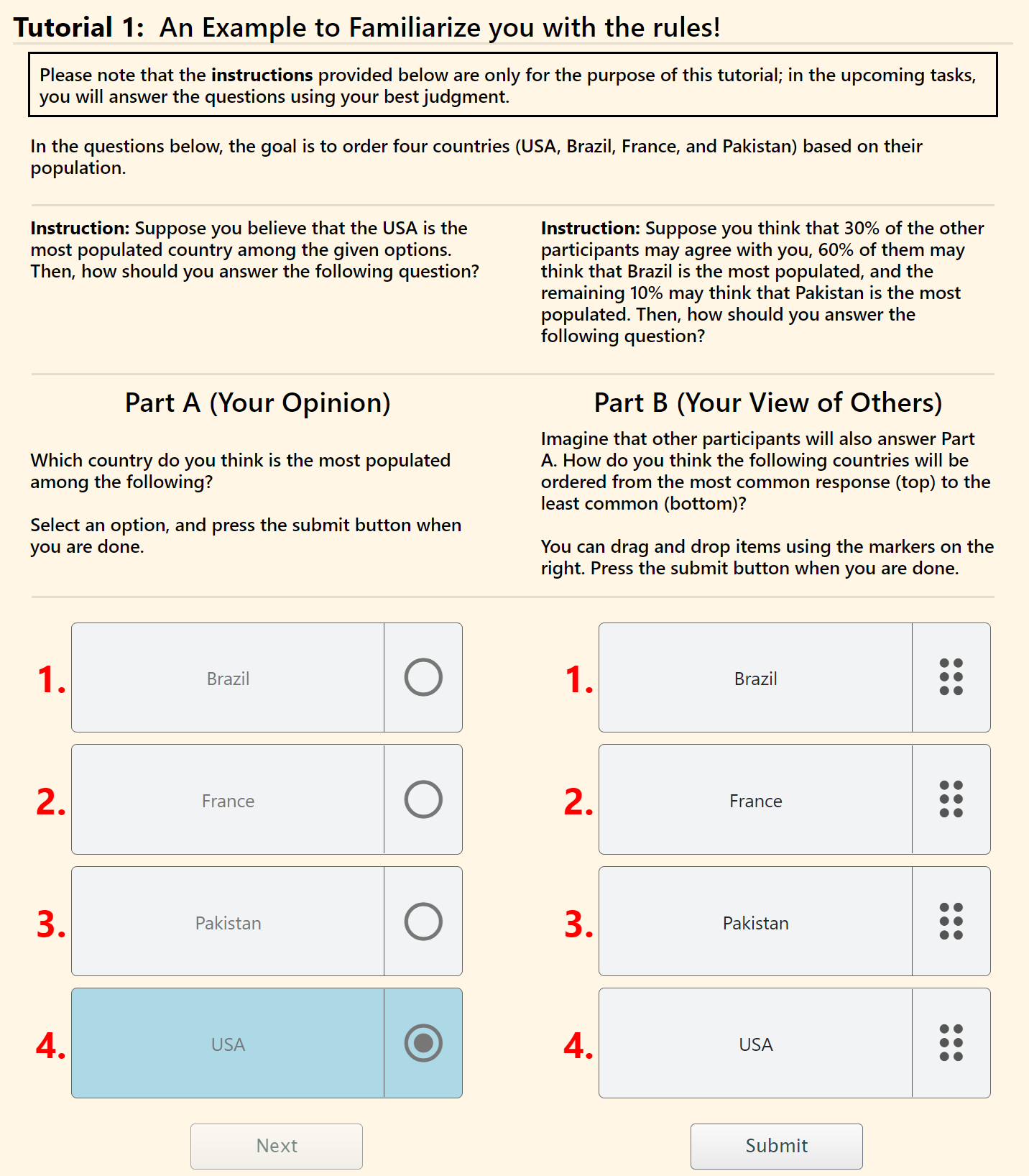}
     \end{subfigure}
        \caption{Tutorials for Rank-Top and Top-Rank formats.  \label{fig:tutorial}}
       
\end{figure*}

Figure \ref{fig:tutorial} shows the tutorials for Rank-Top and Top-Rank elicitation formats. The tutorial provides the turker with a scenario (a correct answer, and a belief about other participants' votes), and asks the turker to complete the vote and prediction (if required) questions. Each turker must successfully complete the tutorial to proceed to the actual questions. 

\newpage

\subsection{Sample Questions}
\begin{figure*}[h!]
     \centering
     \begin{subfigure}[b]{0.32\textwidth}
         \centering
         \includegraphics[width=\textwidth]{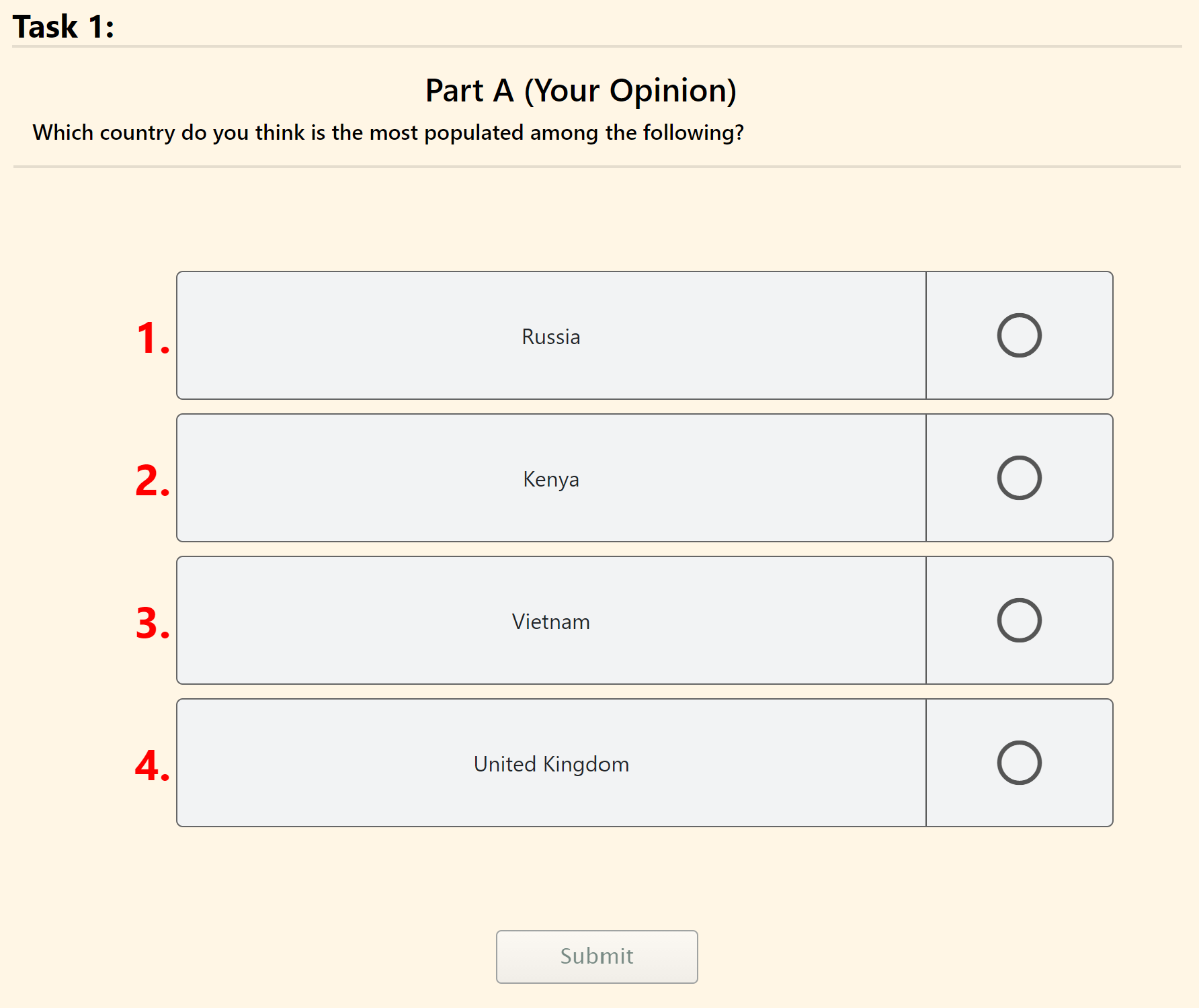}
     \end{subfigure}
     \hfill
     \begin{subfigure}[b]{0.32\textwidth}
         \centering
         \includegraphics[width=\textwidth]{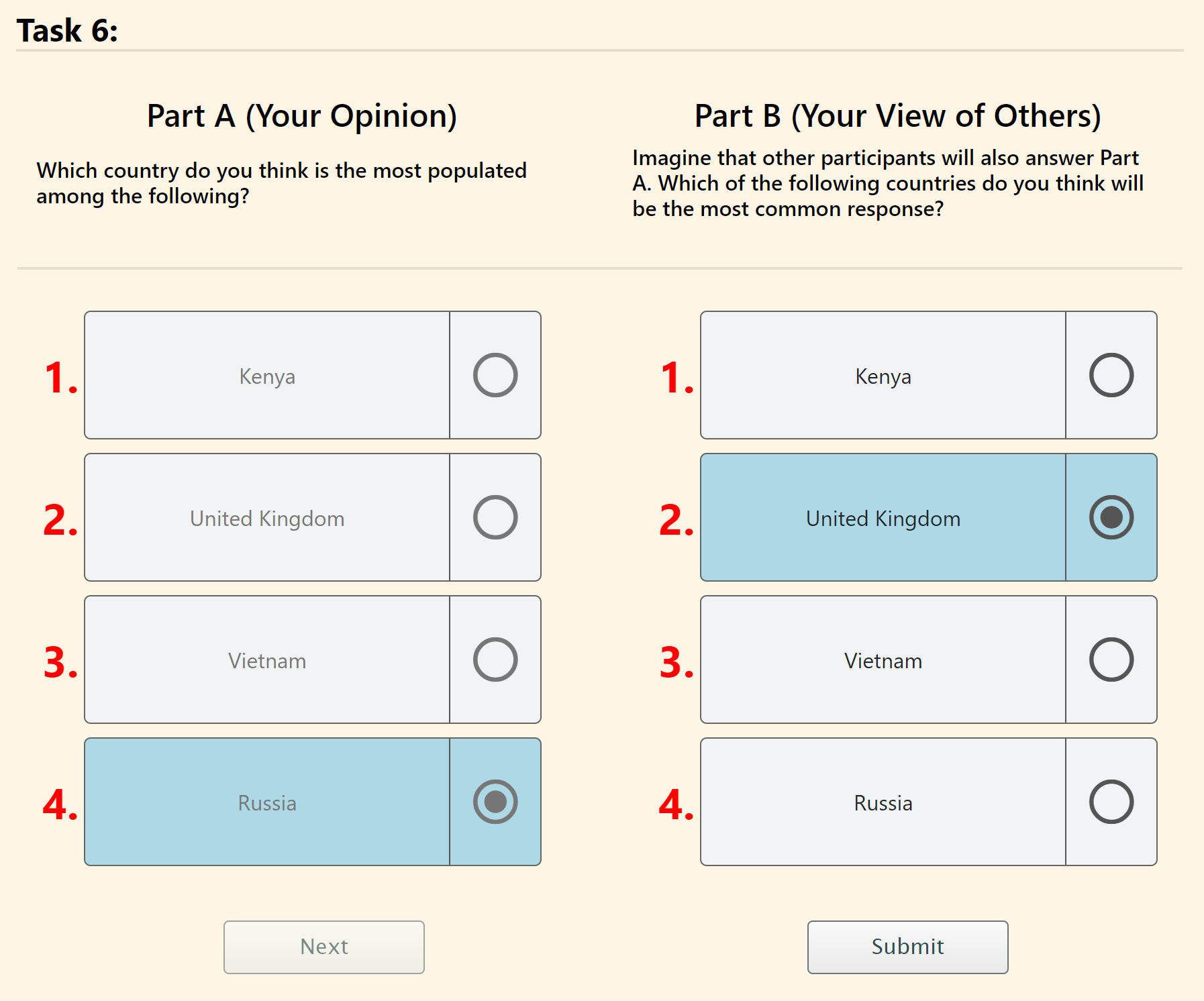}
     \end{subfigure}
     \hfill
    \begin{subfigure}[b]{0.32\textwidth}
         \centering
         \includegraphics[width=\textwidth]{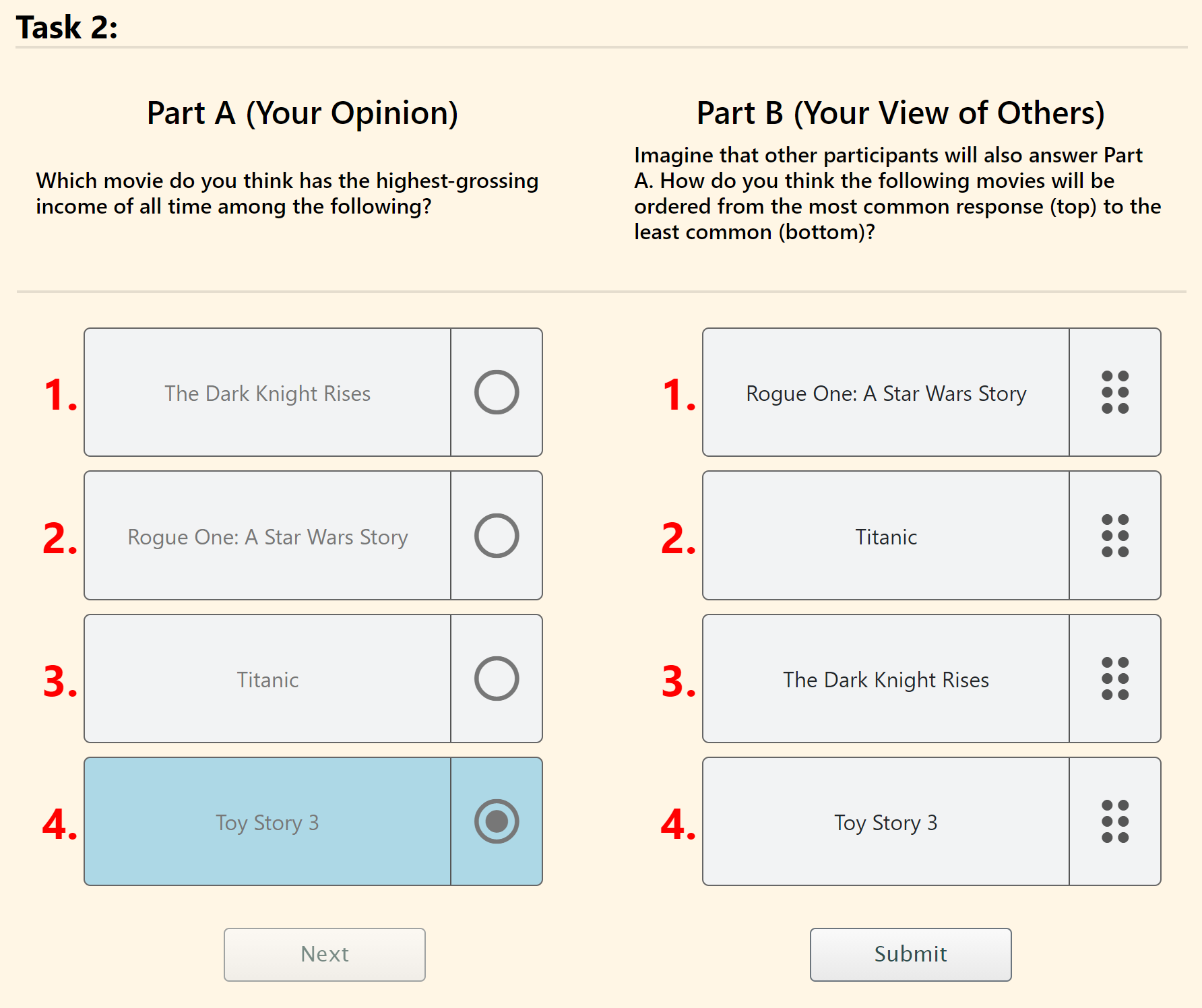}
     \end{subfigure}
        \caption{Sample questions from Top-None, Top-Top, and Top-Rank questions.\label{fig:questions-top}}
\end{figure*}

\begin{figure*}[h!]
     \centering
     \begin{subfigure}[b]{0.31\textwidth}
         \centering
         \includegraphics[width=\textwidth]{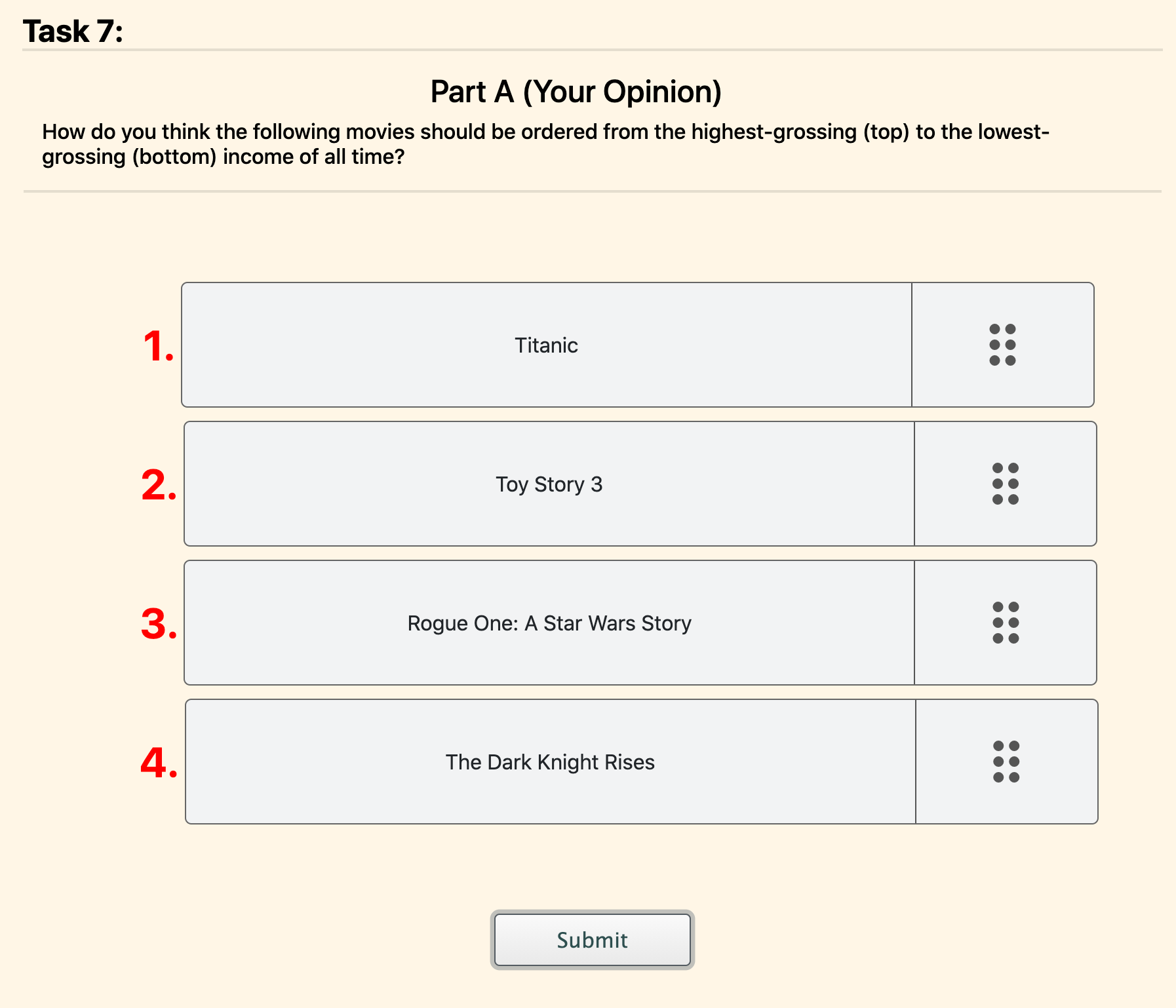}
     \end{subfigure}
     \hfill
     \begin{subfigure}[b]{0.32\textwidth}
         \centering
         \includegraphics[width=\textwidth]{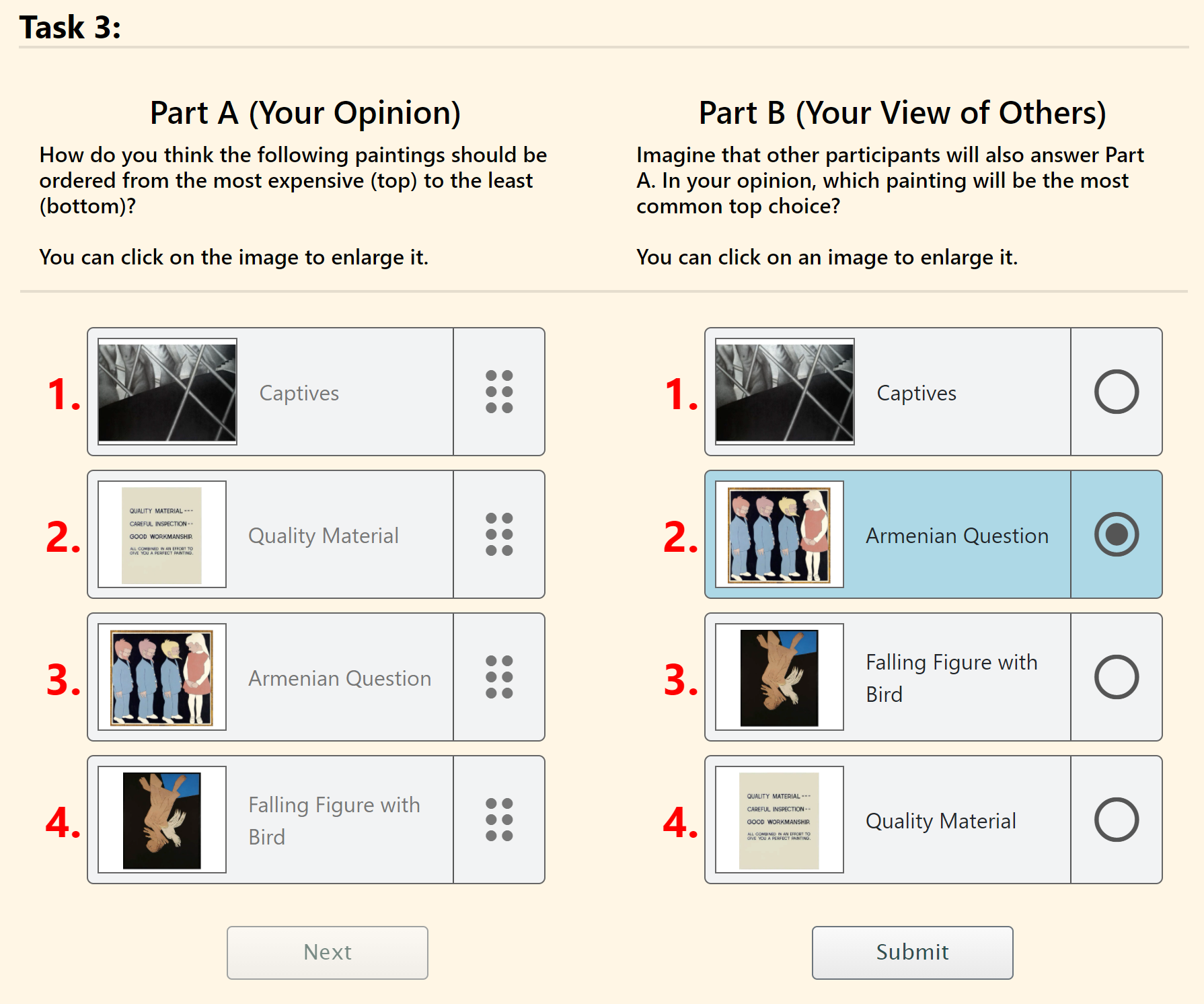}
     \end{subfigure}
     \hfill
    \begin{subfigure}[b]{0.32\textwidth}
         \centering
         \includegraphics[width=\textwidth]{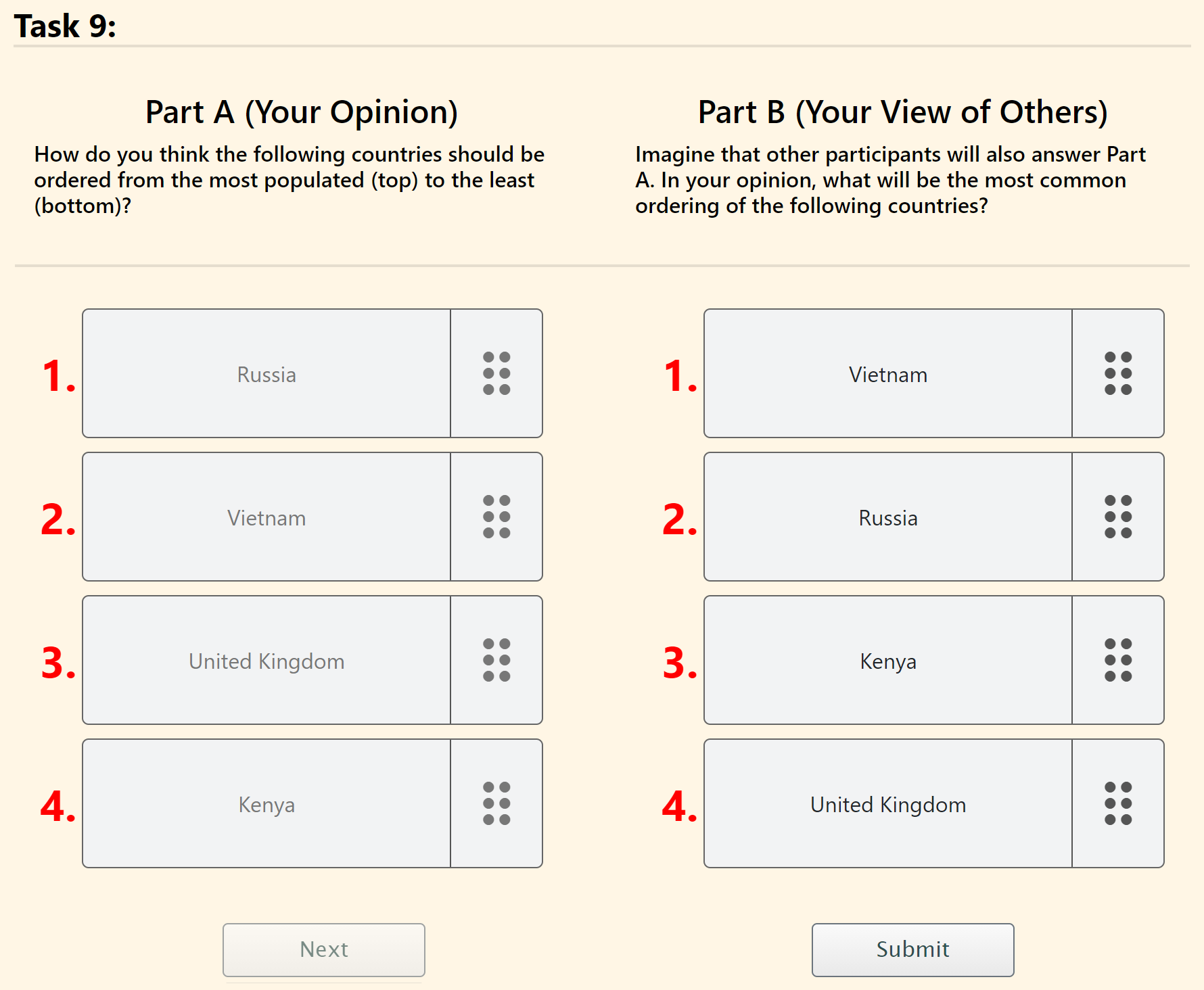}
     \end{subfigure}
        \caption{Sample questions from Rank-None, Rank-Top, and Rank-Rank questions. \label{fig:questions-rank}}
\end{figure*}

Figures \ref{fig:questions-top} and \ref{fig:questions-rank} show sample questions for different elicitation formats. Each turker completes five questions from two elicitation formats.

\subsection{Difficulty/Expressiveness}
\begin{figure*}[h!]
     \centering
     \begin{subfigure}[b]{0.42\textwidth}
         \centering
         \includegraphics[width=\textwidth]{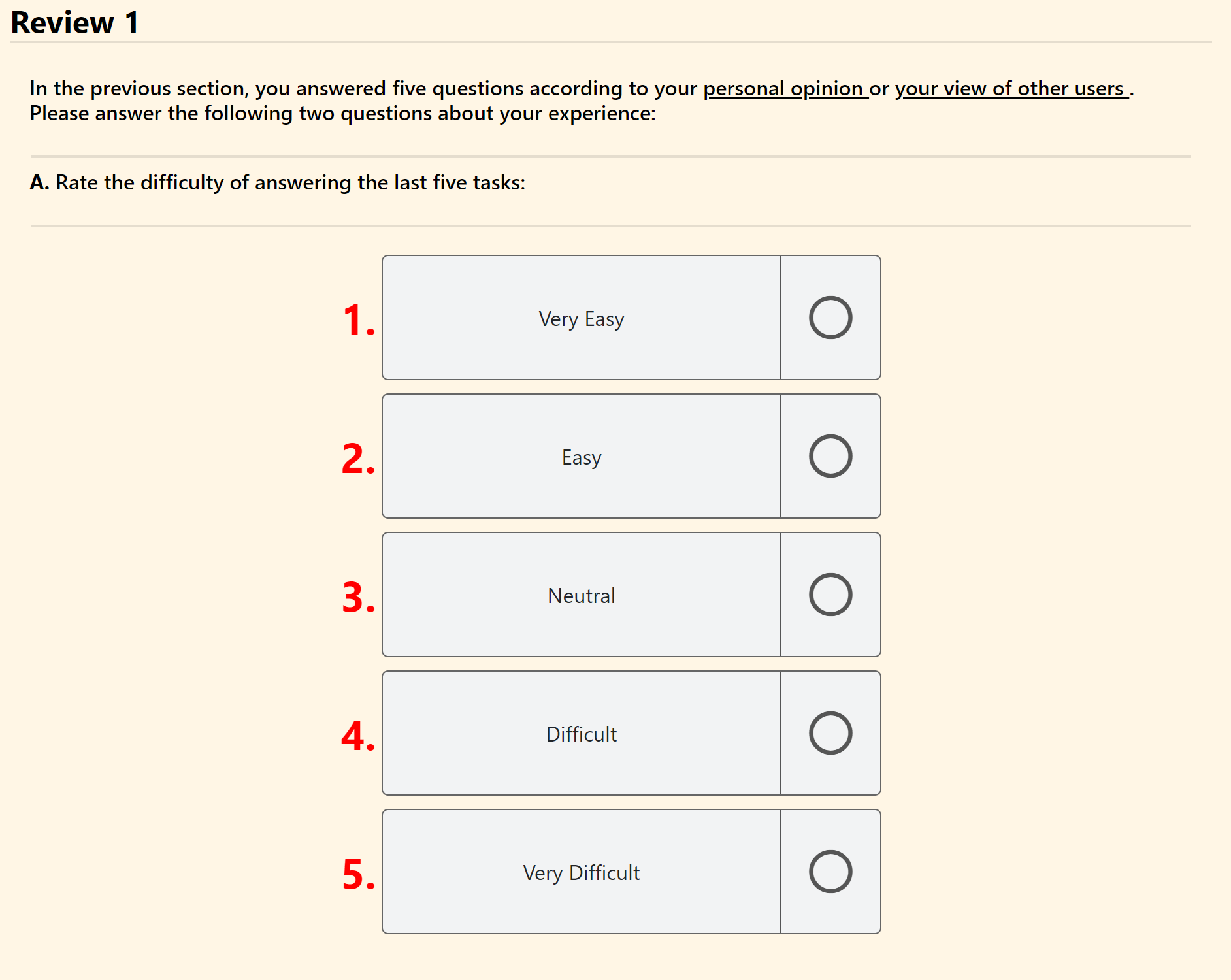}
     \end{subfigure}
     \hfill
     \begin{subfigure}[b]{0.42\textwidth}
         \centering
         \includegraphics[width=\textwidth]{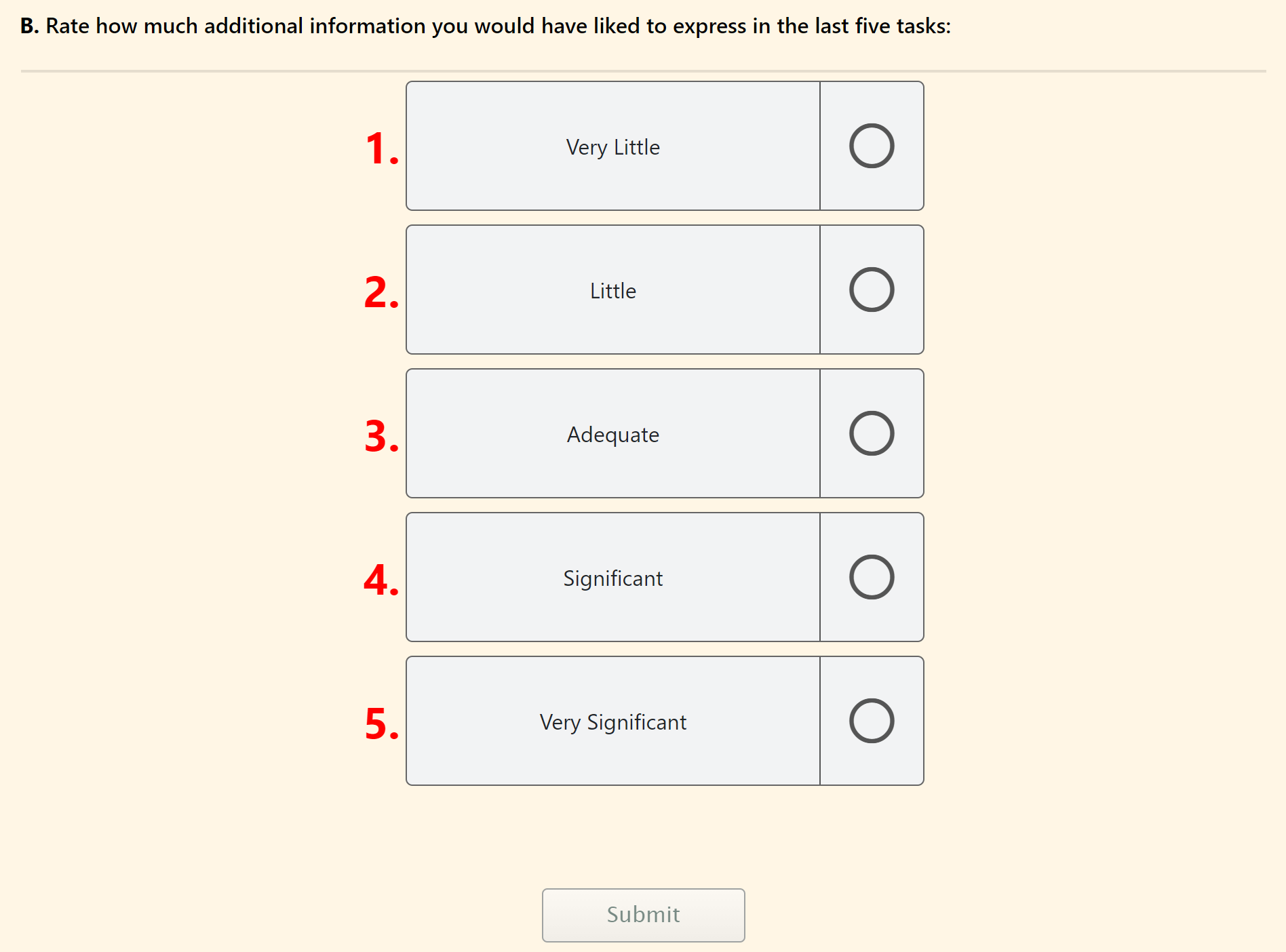}
     \end{subfigure}
        \caption{Difficulty and expressiveness questions \label{fig:dif-expr}}
\end{figure*}

We assign each turker to two elicitation formats. After answering five questions from each format, the turker is asked to complete a survey about the difficulty and expressiveness of that format (shown in figure \ref{fig:dif-expr}).

\end{document}